\documentclass[submitting]{nst}

\usepackage{longtable,array,siunitx}
\usepackage{subfigure,dcolumn}
\usepackage{epstopdf}
\usepackage{mhchem}
\usepackage{upgreek}

\usepackage[]{changes}
\usepackage{extarrows}
\renewcommand\sout{\bgroup \color{red} \ULdepth=-.5ex \ULset}

\begin{document}

\title{Correlation between nuclear isospin asymmetry and $\alpha$-particle preformation probability for superheavy nuclei from a Bayesian inference}
\thanks{Supported by the National Natural Science Foundation of China (Grants Nos: 12175100, 11975132 and 12405154)}

\author{Xiao-Yan Zhu}
\email[Corresponding author, ]{xyzhu0128@163.com}
\affiliation{School of Mathematics and Physics, University of South China, Hengyang, 421001, China}
\affiliation{Hunan Provincial Key Laboratory of Mathematical Modeling and Scientific Computing, University of South China, Hengyang, 421001, China}

\author{Hao Zhang}
\affiliation{School of Nuclear Science and Technology, University of South China, Hengyang, 421001, China}

\author{Wei Gao}
\affiliation{School of Physical Science and Technology, Southwest Jiaotong University, Chengdu, 610031, China}

\author{Wen-Jing Xing}
\email[Corresponding author, ]{wenjing.xing@usc.edu.cn}
\affiliation{School of Nuclear Science and Technology, University of South China, Hengyang, 421001, China}

\author{Xun Chen}
\email[Corresponding author, ]{chenxun@usc.edu.cn}
\affiliation{School of Nuclear Science and Technology, University of South China, Hengyang, 421001, China}

\author{Wen-Bin Lin}
\email[Corresponding author, ]{lwb@usc.edu.cn}
\affiliation{School of Mathematics and Physics, University of South China, Hengyang, 421001, China}
\affiliation{Hunan Provincial Key Laboratory of Mathematical Modeling and Scientific Computing, University of South China, Hengyang, 421001, China}

\author{Xiao-Hua Li}
\email[Corresponding author, ]{lixiaohuaphysics@126.com}
\affiliation{School of Nuclear Science and Technology, University of South China, Hengyang, 421001, China}

\begin{abstract}
In the study of $\alpha$ decay within the superheavy nuclear region ($Z \geq 90$ and $N \geq 140$), the $\alpha$-particle preformation probability $P_{\alpha}$ serves as a crucial physical quantity linking nuclear structure to decay observables. We introduce a phenomenological model that incorporates the decay energy $Q_{\alpha}$, mass number $A$, orbital angular momentum $l$, isospin asymmetry $I$, and the unpaired nucleon effect. For the first time, a Bayesian inference method combined with Markov Chain Monte Carlo (MCMC) sampling has been employed to impose global constraints on the model parameters, enabling the systematic and high-precision calculation of $P_{\alpha}$. The results reveal a significant suppressing effect of isospin asymmetry on $P_{\alpha}$, a finding independently corroborated by random forest-based feature importance analysis, which also identifies $I$ as a feature of non-negligible importance. Furthermore, calculations using the maximum a posteriori (MAP) parameters not only reproduce the shell effect at $N=152$ and 162 but also yield $\alpha$ decay half-life predictions in excellent agreement with experimental ones. This work provides the first global analysis tool within the cosh-type potential (CTP) framework for probing the $\alpha$ preformation mechanism in superheavy nuclei, underscores the potential of the Bayesian framework for inverting complex nuclear physics problems, and establishes a reliable theoretical benchmark for guiding future experimental exploration of superheavy nuclei.
\end{abstract}


\maketitle
\nolinenumbers

\section{Introduction}\label{sec.I}

Since George Gamow \cite{Gamow:1928zz} and Gurrney and Condon \cite{Gurney:1928lxa} first described $\alpha$ decay based on quantum tunneling theory in 1928, quantum mechanics found its initial application in nuclear physics by successfully explaining this process. With the continuous development of radioactive ion beam facilities worldwide, $\alpha$ decay, as one of the dominant decay modes in heavy and superheavy nuclei, has attracted extensive attention \cite{Mang:1964yi,Andreyev:2013iwa,Hamilton:2013una}. Particularly in the synthesis of superheavy elements and the study of nuclear structure, it remains a subject of considerable interest in both experimental and theoretical nuclear physics \cite{Singh:1992hbk,Oganessian:2007zza,Oganessian:2010zz,Ellison:2010zz,ALICE:2012dtf,Ma:2015qga,Hofmann:2000cs}.

Theoretical support plays a crucial role in guiding and interpreting experiments \cite{Wang:2025,Jia2026}. Generally, the primary objective of theoretical studies on $\alpha$ decay is to predict the half-lives and decay modes of unknown nuclei. Various microscopic theories, such as the R-matrix method \cite{Dodig-Crnkovic:1989fpw,Varga:1992zz}, liquid-drop model \cite{Royer:1985mtk,Zhang:2006dj}, Tohsaki-Horiuchi-Schuck-Röpke wave function approach \cite{Ropke:2014wsa,Xu:2015pvv}, and so on \cite{buck1992favoured,Buck:1992zz,Poenaru:2016rbm,Denisov:2005ax,Chowdhury:2005nd,RoyChowdhury:2008uh,Poenaru:2018jju,Zhu:2024swx,Zhu:2025ujz}, have been employed to calculate $\alpha$ decay half-lives. However, a significant factor contributing to the discrepancy between theoretically calculated half-lives and experimental data lies in the determination of the $\alpha$-particle preformation factor, which represents the probability of an $\alpha$-cluster forming on the surface of the parent nucleus. Owing to the structural complexity of quantum many-body systems, accurately calculating the $\alpha$-particle preformation factor, particularly for superheavy nuclei with $Z \geq 90$ and $N \geq 140$, remains exceptionally challenging \cite{Gangopadhyay:2009dke,Guo:2014era,Deng:2020rzy}.

In theoretical calculations, treating the $\alpha$-particle preformation factor of an unknown nucleus as a constant for extrapolating $\alpha$ decay half-lives possesses certain limitations. Consequently, numerous models and phenomenological formulas have been proposed to evaluate $P_{\alpha}$, such as the expression based on the number of valence nucleons (or holes). Building on this approach, studies of nuclei near the $Z=82$ and $N=126$ shell closures have revealed a linear correlation between $P_{\alpha}$ and the product of valence protons (holes) and valence neutrons (holes) \cite{Deng:2017ids,Deng:2018eva,Deng:2021siq}. Moreover, the isospin asymmetry ($I=(N-Z)/A$) serves as a key parameter in nuclear structure and has a significant impact on the $\alpha$ decay process of superheavy nuclei. Theoretical studies have shown that isospin asymmetry not only modulates the nuclear potential and nucleon distribution but also directly suppresses the $\alpha$-particle preformation probability \cite{Seif:2011zz,Ma2025}. For instance, in neutron-rich nuclei, a higher $I$ value may enhance the asymmetry between nucleons, thus reducing the formation efficiency of $\alpha$-cluster. This effect, combined with the number of valence nucleons, makes $I$ a dominant factor in the preformation model. 
Given that the magic numbers in superheavy nuclei remain uncertain, extending the systematics to the superheavy region proves difficult. Subsequent investigations have adopted a microscopic perspective, establishing a connection between $\alpha$ decay energy and $P_{\alpha}$ by empirical half-life formulas, thereby proposing analytical expressions for estimating $P_{\alpha}$. While these phenomenological expressions improve the accuracy of half-life predictions, most of the resulting preformation factors are model-dependent, and their parameters are only locally applicable. In the present work, we adopt the experimental $P_{\alpha}^{\rm{Exp}}$ extracted within the cosh-type potential (CTP) model \cite{Luo:2024ogt} as our benchmark. Therefore, this work aims to develop a global phenomenological model for $P_{\alpha}$ that is consistent with this CTP-based extraction, by incorporating isospin asymmetry explicitly, along with other key factors like $Q_{\alpha}$, $A$, $l$, and unpaired nucleon effects, using a Bayesian inference combined with Markov Chain Monte Carlo (MCMC) sampling to constrain parameters and provide a systematic, high-precision calculation of $P_{\alpha}$ for superheavy nuclei.

The structure of this article is organized as follows. Section \ref{Sec.II} gives theoretical framework. Section \ref{Sec.III} presents the detailed calculations and discussion including the global application of Bayesian inference and $\alpha$-particle preformation factors with isospin effects. Finally, Section \ref{Sec.IV} is a concise summary.

\section{Theoretical framework}\label{Sec.II}
\subsection{$\alpha$-particle preformation factor and local phenomenological expression}
In the CTP model, an analytical phenomenological formula incorporating five parameters has been extracted by combining experimental decay energies and half-lives to represent the $\alpha$-particle preformation factor \cite{Luo:2024ogt}. Utilizing this formula, the preformation factors for nuclides near the neutron magic numbers $N = $126, 152, and 162 were subsequently derived. The results indicate that nuclei in the vicinity of shell closures are more tightly bound compared to their neighboring isotopes.
In this work, $P_{\alpha}$ is extracted from the ratios between theoretical $\alpha$ decay half-life calculated by CTP and the corresponding experimental one. In the framework of CTP, the $\alpha$ decay constant $\lambda$ is defined as 
\begin{equation}
    \lambda=\frac{\hbar P_{\alpha}F P}{4\mu},\label{Eq1}
\end{equation}
where $\hbar$ and $\mu$ represent the Planck constant and reduced mass of the $\alpha$-particle and daughter nucleus. $F$ is the normalized factor of bound-state wave function. $P$ represents the penetration probability calculated using the classical Wentzel-Kramers-Brillouin (WKB) approximation. The experimental $\alpha$ decay constant $\lambda^{Exp}$ can be obtained by
\begin{equation}
    \lambda^{\rm{Exp}}=\frac{\hbar P_{\alpha}^{\rm{Exp}}F P}{4\mu}=\frac{\rm{ln2}}{T_{1/2}^{\rm{Exp}}}.\label{Eq2}
\end{equation}
And assuming the $\alpha$-particle preformation factor to be a constant, $P_{\alpha}=1$, the theoretical $\alpha$ decay constant $\lambda^{\rm{Cal}}$ is calculated by 
\begin{equation}
    \lambda^{\rm{Cal}}=\frac{\hbar P_{\alpha}F P}{4\mu}=\frac{\rm{ln2}}{T_{1/2}^{\rm{Cal}}}.\label{Eq3}
\end{equation}    
The experimental $\alpha$-particle preformation factor can be obtained from the ratio between the theoretical $\alpha$ decay half-life and the corresponding experimental one. It can be expressed as
\begin{equation}
    P_{\alpha}^{\rm{Exp}}=\frac{\lambda^{\rm{Exp}}}{\lambda^{\rm{Cal}}}=\frac{T_{1/2}^{\rm{Cal}}}{T_{1/2}^{\rm{Exp}}}.\label{Eq4}
\end{equation}
In our recent work \cite{Luo:2024ogt}, a local phenomenological expression for estimating $\alpha$-particle preformation factor in heavy and superheavy nuclei has been proposed. It is given by
\begin{equation}
    \mathrm{log}_{10}P_{\alpha}=aZQ_{\alpha}^{-1/2}+bA^{1/3}+c+d[l(l+1)]^{1/2}+h.\label{Eq5}
\end{equation}
Here, parameters $a$, $b$, $c$, $d$, and $h$ are adjustable constants related to physical quantities. 
This expression is constructed similarly to other analytical forms, with the purpose of inferring the possibility of $\alpha$-cluster formation inside the parent nucleus $P_{\alpha}$. This process incorporates all nuclides in the most recent evaluated atomic mass table \cite{Huang:2021nwk, Wang:2021xhn} and is based on the known linear correlations of $Q_{\alpha}$, $Z$, and $A$, combined with an inverse proportionality to the decay half-life. A major drawback of this method, however, is that the adjustable parameters are derived from specific datasets, meaning the resulting expression is only locally applicable.
\subsection{Bayesian inference principle}

Bayesian statistics study the probability of a hypothesis from both current achieved information and previous knowledge. The basis of Bayesian inference is Bayes' rule as
\begin{equation}
    P(\theta|y)=\frac{P(y|\theta)P(\theta)}{P(y)},\label{Bayes}
\end{equation}
where $P(\theta)$ is the prior probability distribution of $\theta$, $P(y|\theta)$ is the likelihood of the data, and $P(y)$ is the marginal likelihood. In this study, a global phenomenological $\alpha$-particle preformation factor model is used. The model employs a multi-dimensional parameter vector $\theta=(\theta_1,\theta_2,...)$. 200 design points are generated by Latin Hypercube Sampling (LHS) \cite{tang1993orthogonal,morris1995exploratory,Foreman-Mackey:2012any}, uniformly distributed across physically meaningful and broadly defined ranges. This forms a $200\times5$ design matrix $\Theta=(\theta_1,\theta_2,...,\theta_{200})^{\top}$. The approach balances coverage of parameter space and computational efficiency.
For efficient Bayesian posterior inference, a Gaussian Process (GP) emulator \cite{mcmillan1999analysis} serves as a surrogate model. It employs an exponentiated quadratic kernel, defined as 
\begin{equation}
   \sigma(\theta,\theta^{'})=\mathrm{exp}(-\frac{||\theta-\theta^{'}||^2}{2t^2}),\label{Eq6}
\end{equation}
where $t$ represents the characteristic length scale that governs the rate of correlation decay between input points. The model also incorporates output centralization. The training output $Y$ is assumed to follow a multivariate Gaussian distribution given by $Y\sim \mathcal{N}(0,K)$, where the covariance matrix $K$ is characterized by elements $[K]_{ij}=\sigma(\theta_i,\theta_j)$. For a new test point $\Theta^{*}$, the GP emulator predicts its output $Y^{*}$ to follow a conditional Gaussian distribution:
\begin{equation}
Y_{*}|\Theta_{*}, \Theta, Y \sim \mathcal{N}(K_{*}K^{-1}Y, K_{**}-K_{*}K^{-1}K_{*}^{\top}).\label{Eq8}
\end{equation}
Here, $K_{*}=\sigma(\Theta_{*}, \Theta)$ denotes the covariance matrix between the new test point and the training data, and $K_{**}=\sigma(\Theta_{*}, \Theta_{*})$ represents the covariance at the test point itself. Although the phenomenological model is analytical and computationally inexpensive for a single parameter set, the Bayesian inference requires extensive sampling of the high-dimensional parameter space via MCMC. Direct evaluation of the model for each sample during MCMC sampling would involve a large number of calculations, which can become costly when iterating over thousands of samples \cite{Jin:2025}. The GP emulator serves as a surrogate model that approximates the model output based on a limited set of training points, allowing for rapid prediction of $\rm{log_{10}}P_{\alpha}$ at new parameter points without recalculating the full model. This significantly accelerates the likelihood evaluation in the MCMC process, improving overall efficiency in parameter calibration and uncertainty quantification.

\section{Results and discussion}\label{Sec.III}

\subsection{The global application of Bayesian inference}

Bayesian inference method has been successfully employed to constrain model parameters across various subfields of physics, including the calibration of relativistic collision models and the analysis of jet energy loss in heavyion physics \cite{He:2018gks,Wu:2023azi,Xing:2023ciw}, as well as studies of the equation of state for nuclear matter \cite{OmanaKuttan:2022aml,Zhu:2025gxo}, nucleon distributions within atomic nuclei \cite{Cheng:2023ucp}, and $\alpha$ decay properties in heavy and superheavy nuclei \cite{Zhu:2025ujz}, among others \cite{Novak:2013bqa,Pratt:2015zsa,Sangaline:2015isa,JETSCAPE:2020shq,Mantysaari:2022ffw}.
In this part, we obtain the global $P_{\alpha}$ by extracting the uncertainties of the input physical quantities given in Eq. (\ref{Eq5}). Provided that different schemes reflect the essential feature of $P_{\alpha}$ and yield results that are consistent with experimental data after Bayesian calibration, a global analytical expression can be established. 

In the following, we employ the Bayesian inference approach, leveraging experimental data on \replaced{164}{} superheavy nuclei with $Z \geq 90$ and $N \geq 140$ from Ref. \cite{Zhu:2025ujz}, to calibrate and quantify uncertainties in the parameters $\theta=(a, b, c, d, h)$ of Eq. (\ref{Eq5}). This calibration process constitutes an inverse problem, wherein the model input parameters are inferred from experimental data $y^{\rm exp}$. The corresponding statistical inference is grounded in the posterior distribution, which is expressed as
\begin{equation}
P(\theta|y^{\rm exp}) \varpropto P(y^{\rm exp}|\theta) \cdot P(\theta),\label{Eq9}
\end{equation}
where $P(\theta)$ is the prior distribution. We employ uniform priors over physically plausible parameter ranges $a\in(-0.15,0.05)$, $b\in(-2.0,2.0)$, $c\in(4.0,16.0)$, $d \in(-0.25,0.67)$, $h\in(-0.8,-0.1)$, which are sufficiently broad to allow the model outputs to cover all experimental data. The likelihood function $P(y^{\rm exp}|\theta)$ is formulated based on a Gaussian assumption and expressed in a $\chi^{2}$ form:
\begin{equation}
    P(\mathrm{y^{\rm exp}}|\theta)= \prod_{i} \frac{1}{\sqrt{2\pi}\sigma_i}e^{-(y_i(\theta)-y_i^{\mathrm{exp}})^2/(2\sigma_i^2)},\label{Eq9}
\end{equation}
where $y_i(\theta)$ denotes the predicted value of the model at the $i$-th data point. $y_i^{\mathrm{exp}}$ and $\sigma_i$ represent the corresponding experimental data and their associated uncertainty, respectively. To conduct a comprehensive assessment of the consistency between the parameters and the entire set of experimental data, a joint likelihood function is constructed by integrating the data from even-even (ee), odd-A (oa), and odd-odd (oo) nuclei. It is given by
\begin{equation}
\begin{array}{l}
    P(y^{\rm exp}|\theta)= P(y^{\rm exp}_{\mathrm{log_{10}P_{\alpha}^{ee}}}|\theta)
    P(y^{\rm exp}_{\mathrm{log_{10}P_{\alpha}^{oa}}}|\theta)P(y^{\rm exp}_{\mathrm{log_{10}P_{\alpha}^{oo}}}|\theta) .\label{Eq.10}
\end{array}
\end{equation}

To effectively sample from the posterior distribution, we employ the Metropolis-Hastings algorithm to conduct the MCMC method \cite{Goodman:2010dyf,Foreman-Mackey:2012any} in the parameter space. 
The sampling process is performed in logarithmic probability space to enhance numerical stability, with new parameter proposals being generated by a Gaussian transition kernel centered on the current position. To further improve sampling efficiency, an ensemble of 150 walkers is initialized from random starting points within the parameter space. Following a burn-in phase of 5,000 steps to ensure the chains have adequately converged to the target posterior distribution, a subsequent set of 10,000 steps per walker is recorded to generate the final posterior sample ensemble.

\begin{figure}[!htb]\centering
 \includegraphics
  [width=0.8\hsize]
  {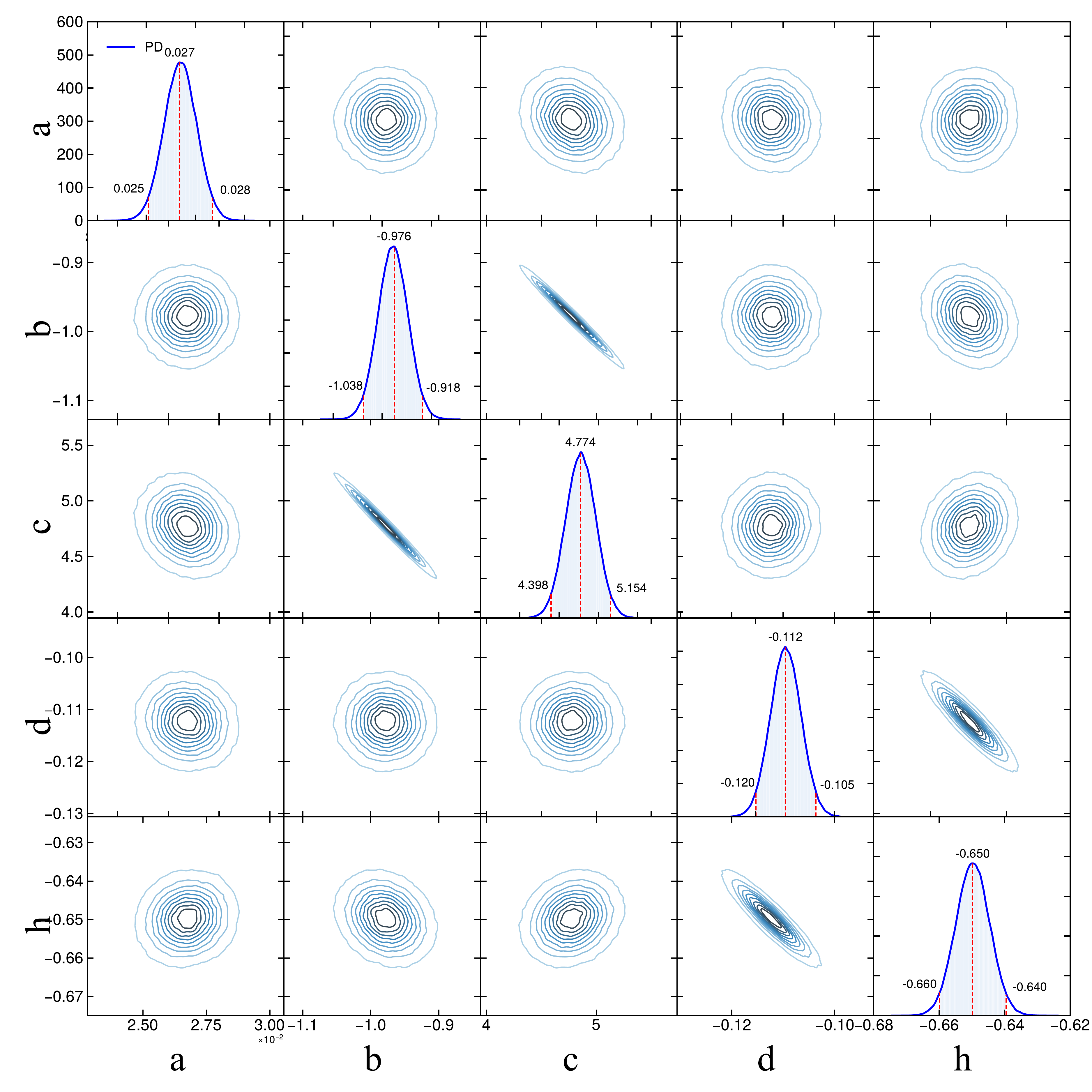}
 \caption{Posterior distributions of the model's global parameters (diagonal panels) and their
correlations (off-diagonal panels) extracted from Eq. (\ref{Eq5}) using the original uncertainties of the experimental data.}
 \label{Fig1}
 \end{figure}

Based on the MCMC sampling results, the marginal posterior distributions of the parameters are estimated, with their maximum a posteriori (MAP) estimates and $95\%$ credible intervals (C.I.'s) obtained. 
The correlations and degeneracies among the parameters are visualized using a corner plot, as illustrated in Fig. \ref{Fig1}. The diagonal subpanels of this figure display marginal posterior distributions for each parameter, where the MAP estimates and $95\%$ C.I.'s are marked by red dashed lines. The off-diagonal subpanels, conversely, present the two-dimensional joint posterior distributions for each parameter pair. As can be seen from Fig. \ref{Fig1}, the posterior distribution of parameter $a$ (associated with $ZQ_{\alpha}^{-1/2}$) is concentrated and independent, suggesting a well-constrained and stable underlying physical mechanism. In contrast, the strong positive correlations observed among parameters $b, c, d,$ and $h$ indicate potential coupling or redundancy in the physical effects related to mass number $A^{1/3}$ and angular momentum $(l(l+1))^{1/2}$ within the model. This correlation structure provides crucial clues for understanding the microscopic mechanism of the $\alpha$-particle preformation factor.

\subsection{$\alpha$-particle preformation factors with isospin effects}
For an extended period, decay has been regarded as a reliable avenue for investigating nuclear structural information. Among various decay modes, $\alpha$ decay, which stands as one of the predominant decay mechanisms in heavy and superheavy nuclei, continues to serve as one of the most crucial and effective means for probing the structure, properties, and synthesis mechanisms of superheavy nuclei \cite{Andreyev:2013iwa,Ma:2015qga,Hofmann:2000cs}.
Recent studies \cite{Shin:2015iiw,Saxena:2024vtg} have revealed that the isospin asymmetry effect plays an essential role in the $\alpha$ decay life-time and the key physical quantity $P_{\alpha}$, by simultaneously influencing the nuclear potential and proton distribution. Based on this understanding, the present work focuses on investigating the specific manifestations of the isospin asymmetry effect in the $\alpha$-particle preformation factor within the superheavy nuclear region ($Z \geq 90$ and $N \geq 140$). To construct a global phenomenological model for $P_{\alpha}$, we employ the same experimental data and Bayesian inference method as described in the preceding section. The model adopts $Z$, $Q_{\alpha}^{-1/2}$, $A^{1/3}$, $l$, and the isospin asymmetry parameter $I=(N-Z)/A$ as feature variable, with $\mathrm{log}_{10}P_{\alpha}$ as the target variables. It is given by
\begin{equation}
    \mathrm{log}_{10}P_{\alpha}=aZQ_{\alpha}^{-1/2}+bA^{1/3}+c+d[l(l+1)]^{1/2}+e[I(I+1)]^{1/2}+h.\label{Eq12}
\end{equation}
Here, parameter $a$, associated with $Q_{\alpha}$ and $Z$, quantifies the contribution of decay energy to $\mathrm{log}_{10}P_{\alpha}$. Parameter $b$, which relates to $A^{1/3}$, reflects the modulation of nuclear surface effects. The constant term $c$ represents other unspecified systematic effects or serves as a baseline value. Parameter $d$, connected with $l$, characterizes the hindrance effect of the centrifugal barrier on $P_{\alpha}$.
Parameter $e$, associated with the neutron-proton asymmetry $I$, describes the influence of the isospin effect on $P_{\alpha}$. Lastly, Parameter $h$, linked to unpaired nucleons, accounts for the blocking effect resulting from nucleonic pairing correlations.
\begin{figure*}[!htbp]
\centering
 \includegraphics
  [width=0.8\hsize]
  {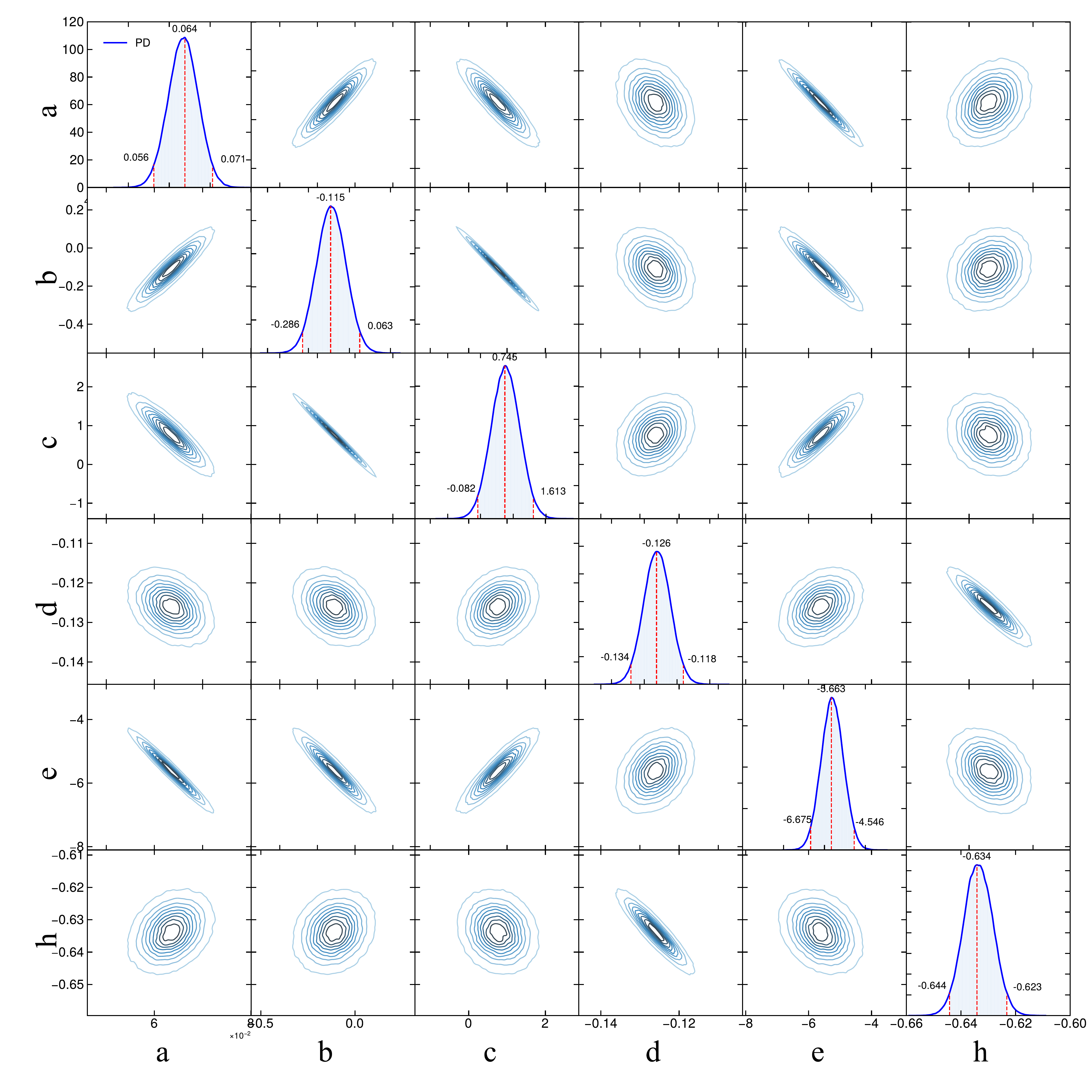}
 \caption{Posterior distributions of the model's global parameters and their
correlations extracted from Eq. (\ref{Eq12}) using the original uncertainties of the experimental data.}
 \label{Fig2}
 \end{figure*}

In the model considering isospin effects, Eq. (\ref{Eq12}), the parameter $\theta=(a, b, c, d, e, h)$ is a six-dimensional vector. A uniform prior distribution $P(\theta)$ is assumed over the following intervals: $a\in(0.02,0.1)$, $b\in(-2.0,2.0)$, $c\in(-2.0,4.0)$, $d\in(-0.3,0.1)$, $e\in(-10.0,-2.0)$, and $h\in(-1.0,-0.2)$. To improve the efficiency of parameter calibration, \replaced{a}{} GP emulator is employed to scan the six-dimensional parameter space, serving as a surrogate for the actual perturbative calculations. This approach accelerates the evaluation of the likelihood in the Bayesian analysis. Using the same MCMC sampling method as described in the previous section, Fig. \ref{Fig2} displays the posterior probability distributions of the six parameters obtained from Bayesian inference to the experimental data. Along the diagonal are the marginal posterior distributions of individual parameters, with red dashed lines indicating MAP and $95\%$ C.I.'s. The off-diagonal subplots display the joint posterior distributions between parameters. From Fig. \ref{Fig2}, physically meaningful relationships among parameters can be observed. Parameter $a$ exhibits a markedly positive peak value, indicating that $Q_{\alpha}$ serves as the primary enhancer of $\alpha$ decay by increasing the penetration probability $P$ through the barrier-lowering effect. In contrast, the $Q_{\alpha}^{-1/2}$term in Eq. (\ref{Eq12}) contributes to a reduction in the preformation probability $P_{\alpha}$, reflecting the indirect influence of decay energy on cluster formation due to factors such as isospin asymmetry. The posterior distribution of parameter $b$ is concentrated in the negative region, implying that the size effect represented by $A^{1/3}$ significantly suppresses the preformation probability of $\alpha$-particle in the superheavy region, consistent with the higher Coulomb barrier and more complex cluster formation process in superheavy nuclei. Parameter $e$ is negative and possesses a large absolute magnitude, providing strong evidence that the isospin effect inhibits $\alpha$ decay, reflecting the considerable difficulty of forming an $N=Z$ $\alpha$-cluster in extremely neutron-rich superheavy nuclei. The MAP value $h$ is also negative, suggesting that unpaired nucleons may impose additional suppression on $P_{\alpha}$. It is evident that in superheavy nuclei, the preformation probability of $\alpha$ decay is modulated collectively by the enhancing effect of $Q_{\alpha}$, the suppressing influence of $A^{1/3}$, the inhibition due to strong neutron–proton asymmetry, and possible shell effects. Among these, the distinctly negative value of parameter $e$ and its relative independence observed in the joint distributions underscore the necessity of incorporating an isospin asymmetry correction term when describing $\alpha$ decay in superheavy nuclei.
\begin{figure*}[!htbp]\centering
 \includegraphics
  [width=0.7\hsize]
  {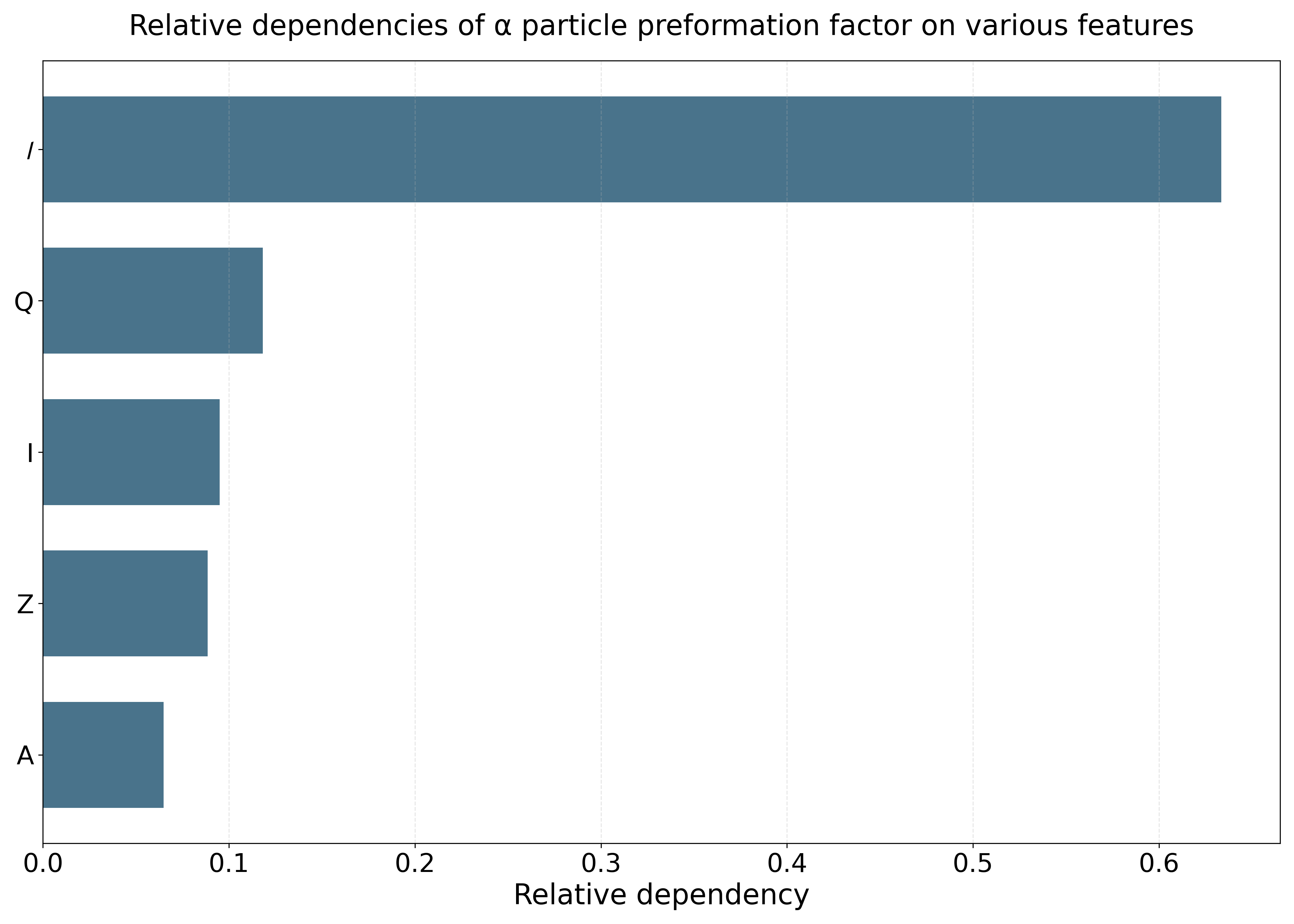}
 \caption{Relative dependencies of $\alpha$ particle preformation factor on various features.}
 \label{Fig3}
 \end{figure*}

Furthermore, we employ the random forest method to evaluate the relative importance of the features influencing the $\alpha$-particle preformation factor. The random forest method is an ensemble machine learning algorithm that constructs multiple decision trees. Feature importance is quantified by the mean decrease in impurity when a given feature is used for splitting across different trees, with a higher numerical value indicating a stronger influence. This approach provides independent validation for dominant factors such as isospin asymmetry, analyzing their relative dependencies without presuming the model structure. As shown in Fig. \ref{Fig3}, the results reveal that, among the selected features $(l, Q, I , Z, A)$, the relative dependence of $I$ is relatively high, exceeding that of $Z$ and $A$. Although the feature importance ranking does not directly reflect the magnitudes or signs of the parameters $(a, b, c, d, e, h)$, this machine learning analysis provides an independent, model-agnostic perspective on the input variables. The random forest analysis highlights the overall predictive role of each variable in the dataset, while Bayesian inference quantifies the specific inhibitory effect of item $I$ included in our model.

\begin{figure*}[!htbp]\centering
 \includegraphics
  [width=0.8\hsize]
  {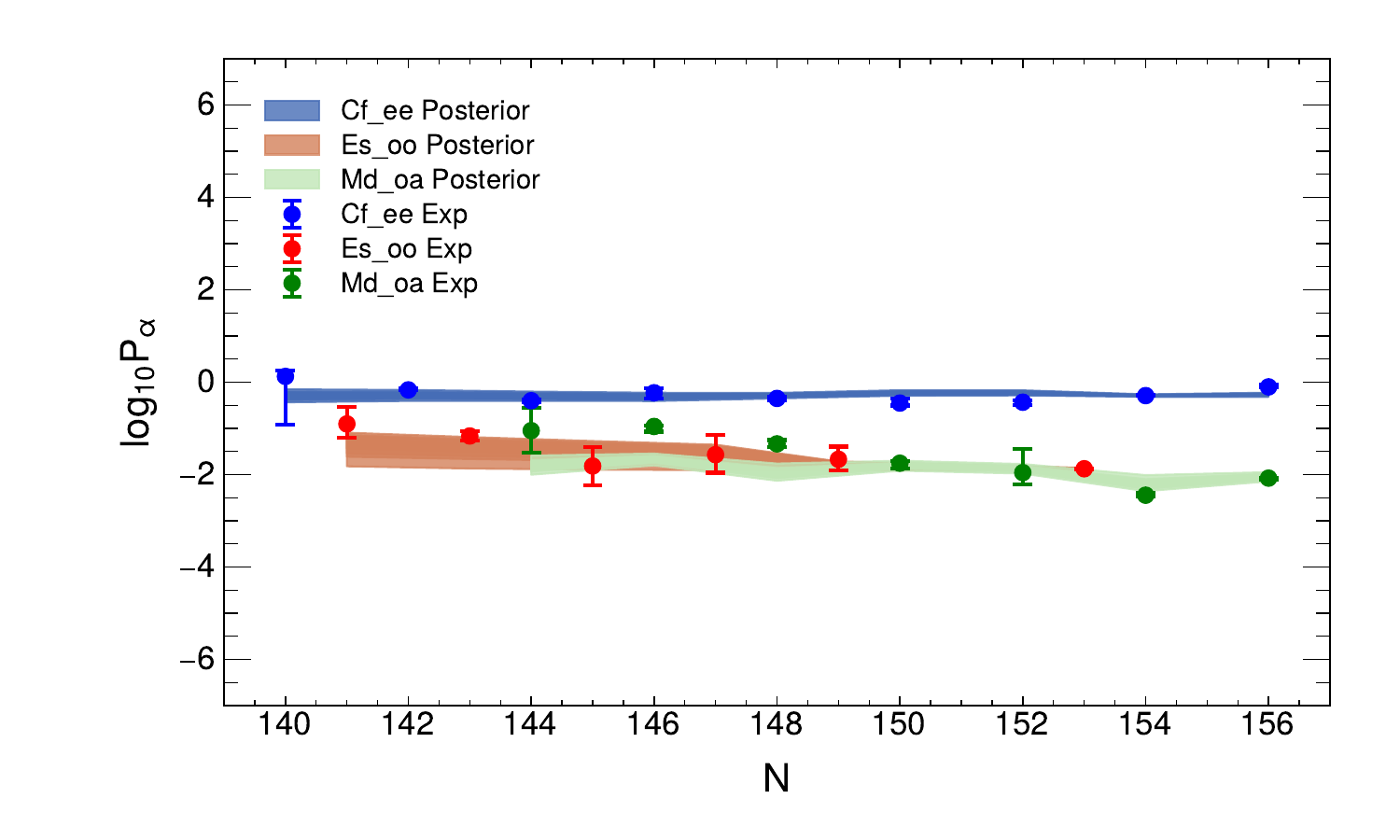}
 \caption{The calculation results of the posterior distribution $P_{\alpha}$ of the parameters of for Cf, Es, and Md nuclei based on the model Eq. (\ref{Eq12}) are compared with the experimental data.}
 \label{Fig4}
 \end{figure*}
 
Table \ref{table1} provides data on 164 superheavy nuclei with $Z \geq 90$ and $N \geq 140$. The first column lists the decaying nuclei, followed by the experimental decay energies with uncertainties, orbital angular momenta, and the isospin asymmetry parameter. The fifth column presents the logarithm of the experimental half-lives, also with uncertainties. Columns six and seven show the calculated preformation factors, denoted as $\rm{P_{\alpha}^{{Cal1}}}$ and $\rm{P_{\alpha}^{{Cal2}}}$, which are obtained by Eq. (\ref{Eq5}) and Eq. (\ref{Eq12}) using the respective MAP parameters, respectively. Subsequent columns eight and nine present the corresponding calculated $\alpha$ decay half-lives, labeled as ${\rm{log}}_{10}T_{1/2}^{\rm{{Cal1}}}$ and ${\rm{log}}_{10}T_{1/2}^{\rm{{Cal2}}}$. A closer agreement is observed between the ${\rm{log}}_{10}T_{1/2}^{\rm{{Cal2}}}$ values, calculated using (\ref{Eq12}), and the experimental data. In this model, the explicit contribution of isospin asymmetry to the preformation probability of $\alpha$-particle is negative, indicating an inhibitory effect on $\rm{P_{\alpha}}$. However, the overall trend of $\rm{P_{\alpha}}$ is influenced by multiple factors, and in some nuclides, it may exhibit different behaviors due to other physical quantities. This outcome is further illustrated in Fig. \ref{Fig4}, which presents the calculated $\alpha$-particle preformation factors based on the Bayesian calibrated model. Using the MAP values of parameters drawn from the posterior distribution, the preformation factors are calculated and compared with experimental ones for even-even (Cf), odd-even (Es), and odd-A (Md) nuclei. The posterior distributions derive from the calibrated model (depicted as blue, brown, and green bands) show good agreement with the corresponding experimental ones (including error bars) across most neutron number intervals, with the most consistent description achieved for the even-even Cf nuclei. In Fig. \ref{Fig4}, the posterior distributions for $\rm{P_{\alpha}}$ are depicted as bands, which represent the $95\%$ Bayesian credible intervals derived from the MCMC samples. These intervals quantify the uncertainty in the preformation factors, while the comparisons are made with experimental data points including error bars. The results are based on the posterior mean values for clarity.

\renewcommand\arraystretch{0.8}
\setlength{\tabcolsep}{1.2mm}
\begin{longtable*}{ccccccccc}
	\caption{The $\alpha$-particle preformation factors and half-lives of \replaced{164}{} superheavy nuclei with $Z \geq 90$ and $N \geq 140$ are calculated using Eq. (\ref{Eq5})and Eq. (\ref{Eq12}). The experimental $\alpha$ decay energies and half-lives are taken from the Refs. \cite{Zhu:2024swx, Huang:2021nwk, Wang:2021xhn}.}\\
	\hline
    \toprule
	$\alpha$ transition & $\rm{Q_{\alpha}^{Exp}}$&$l$ &I    &${\rm{log}}_{10}T_{1/2}^{\rm{{Exp}}}$&$\rm{P_{\alpha}^{{Cal1}}}$&$\rm{P_{\alpha}^{{Cal2}}}$
         &${\rm{log}}_{10}T_{1/2}^{\rm{{Cal1}}}$  &${\rm{log}}_{10}T_{1/2}^{\rm{{Cal2}}}$\\
         \hline
	\endfirsthead
	\multicolumn{8}{c}
	{{\tablename\ \thetable{} -- continued from previous page}} \\
	\hline
	\endhead
	\hline \multicolumn{8}{r}{{Continued on next page}} \\
	\endfoot
	\endlastfoot
$	^{230}	$Th$	\to	^{226}	$Ra$	$	&	4.7700 	$\pm$	0.0015 	&	0	&	0.2174 	&	$	12.3765 	^{+	0.0017 	}_{-	0.0017 	}	$	&	0.4949 	&	0.4354 	&	12.7209 	&	12.7765 	\\	
$	^{232}	$Th$	\to	^{228}	$Ra$	$	&	4.0816 	$\pm$	0.0014 	&	0	&	0.2241 	&	$	18.6452 	^{+	0.0031 	}_{-	0.0031 	}	$	&	0.5802 	&	0.6222 	&	18.0568 	&	18.0265 	\\	
$	^{231}	$Pa$	\to	^{227}	$Ac$	$	&	5.1499 	$\pm$	0.0008 	&	0	&	0.2121 	&	$	12.0130 	^{+	0.0027 	}_{-	0.0027 	}	$	&	0.1009 	&	0.0933 	&	11.4346 	&	11.4686 	\\	
$	^{232}	$U$	\to	^{228}	$Th$	$	&	5.4136 	$\pm$	0.0001 	&	0	&	0.2069 	&	$	9.3373 	^{+	0.0025 	}_{-	0.0025 	}	$	&	0.4300 	&	0.4125 	&	9.7721 	&	9.7902 	\\	
$	^{233}	$U$	\to	^{229}	$Th$	$	&	4.9087 	$\pm$	0.0012 	&	0	&	0.2103 	&	$	12.7010 	^{+	0.0004 	}_{-	0.0004 	}	$	&	0.1055 	&	0.1180 	&	13.5277 	&	13.4789 	\\	
$	^{234}	$U$	\to	^{230}	$Th$	$	&	4.8575 	$\pm$	0.0007 	&	0	&	0.2137 	&	$	12.8891 	^{+	0.0011 	}_{-	0.0011 	}	$	&	0.4711 	&	0.4987 	&	13.2185 	&	13.1938 	\\	
$	^{236}	$U$	\to	^{232}	$Th$	$	&	4.5730 	$\pm$	0.0009 	&	0	&	0.2203 	&	$	14.8687 	^{+	0.0007 	}_{-	0.0007 	}	$	&	0.4884 	&	0.5290 	&	15.2425 	&	15.2078 	\\	
$	^{238}	$U$	\to	^{234}	$Th$	$	&	4.2699 	$\pm$	0.0021 	&	0	&	0.2269 	&	$	18.1487 	^{+	0.0003 	}_{-	0.0003 	}	$	&	0.5134 	&	0.5814 	&	17.6283 	&	17.5743 	\\	
$	^{233}	$Np$	\to	^{229}	$Pa$	$	&	5.6300 	$\pm$	0.0500 	&	0	&	0.2017 	&	$	8.4918 	^{+	0.0012 	}_{-	0.0012 	}	$	&	0.0918 	&	0.0987 	&	9.6738 	&	9.6421 	\\	
$	^{235}	$Np$	\to	^{231}	$Pa$	$	&	5.1938 	$\pm$	0.0015 	&	1	&	0.2085 	&	$	12.1193 	^{+	0.0013 	}_{-	0.0013 	}	$	&	0.0670 	&	0.0721 	&	12.4516 	&	12.4193 	\\	
$	^{236}	$Np$	\to	^{232}	$Pa$	$	&	5.0100 	$\pm$	0.0500 	&	4	&	0.2119 	&	$	15.4797 	^{+	0.0140 	}_{-	0.0144 	}	$	&	0.0068 	&	0.0070 	&	15.3687 	&	15.3579 	\\	
$	^{237}	$Np$	\to	^{233}	$Pa$	$	&	4.9573 	$\pm$	0.0007 	&	1	&	0.2152 	&	$	13.8303 	^{+	0.0014 	}_{-	0.0014 	}	$	&	0.0681 	&	0.0730 	&	13.9700 	&	13.9400 	\\	
$	^{234}	$Pu$	\to	^{230}	$U$	$	&	6.3100 	$\pm$	0.0050 	&	0	&	0.1966 	&	$	5.7226 	^{+	0.0049 	}_{-	0.0050 	}	$	&	0.3637 	&	0.3658 	&	6.0213 	&	6.0187 	\\	
$	^{235}	$Pu$	\to	^{231}	$U$	$	&	5.9510 	$\pm$	0.0200 	&	0	&	0.2000 	&	$	7.7341 	^{+	0.0085 	}_{-	0.0087 	}	$	&	0.0847 	&	0.0921 	&	8.4243 	&	8.3877 	\\	
$	^{236}	$Pu$	\to	^{232}	$U$	$	&	5.8672 	$\pm$	0.0001 	&	0	&	0.2034 	&	$	7.9552 	^{+	0.0012 	}_{-	0.0012 	}	$	&	0.3795 	&	0.3919 	&	8.2026 	&	8.1886 	\\	
$	^{238}	$Pu$	\to	^{234}	$U$	$	&	5.5933 	$\pm$	0.0002 	&	0	&	0.2101 	&	$	9.4421 	^{+	0.0005 	}_{-	0.0005 	}	$	&	0.3857 	&	0.3949 	&	9.6865 	&	9.6763 	\\	
$	^{239}	$Pu$	\to	^{235}	$U$^{\rm{m}}$$	$	&	5.2445 	$\pm$	0.0002 	&	0	&	0.2134 	&	$	11.8813 	^{+	0.0005 	}_{-	0.0005 	}	$	&	0.0909 	&	0.1026 	&	12.4078 	&	12.3552 	\\	
$	^{240}	$Pu$	\to	^{236}	$U$	$	&	5.2558 	$\pm$	0.0001 	&	0	&	0.2167 	&	$	11.3161 	^{+	0.0005 	}_{-	0.0005 	}	$	&	0.3996 	&	0.4179 	&	11.6478 	&	11.6283 	\\	
$	^{241}	$Pu$	\to	^{237}	$U$	$	&	5.1401 	$\pm$	0.0500 	&	2	&	0.2199 	&	$	13.2661 	^{+	0.0009 	}_{-	0.0009 	}	$	&	0.0471 	&	0.0467 	&	13.6435 	&	13.6464 	\\	
$	^{242}	$Pu$	\to	^{238}	$U$	$	&	4.9842 	$\pm$	0.0010 	&	0	&	0.2231 	&	$	13.0731 	^{+	0.0023 	}_{-	0.0023 	}	$	&	0.4102 	&	0.4339 	&	13.3818 	&	13.3575 	\\	
$	^{244}	$Pu$	\to	^{240}	$U$	$	&	4.6656 	$\pm$	0.0010 	&	0	&	0.2295 	&	$	15.4097 	^{+	0.0016 	}_{-	0.0016 	}	$	&	0.4292 	&	0.4727 	&	15.6506 	&	15.6087 	\\	
$	^{235}	$Am$	\to	^{231}	$Np$	$	&	6.5760 	$\pm$	0.0130 	&	1	&	0.1915 	&	$	5.1889 	^{+	0.0246 	}_{-	0.0261 	}	$	&	0.0535 	&	0.0574 	&	6.1683 	&	6.1370 	\\	
$	^{236}	$Am$	\to	^{232}	$Np$	$	&	6.2600 	$\pm$	0.0600 	&	0	&	0.1949 	&	$	6.7324 	^{+	0.0119 	}_{-	0.0122 	}	$	&	0.0178 	&	0.0213 	&	8.0326 	&	7.9548 	\\	
$	^{237}	$Am$	\to	^{233}	$Np$	$	&	6.2000 	$\pm$	0.0300 	&	1	&	0.1983 	&	$	7.2471 	^{+	0.0047 	}_{-	0.0047 	}	$	&	0.0549 	&	0.0590 	&	7.9008 	&	7.8690 	\\	
$	^{239}	$Am$	\to	^{235}	$Np$	$	&	5.9224 	$\pm$	0.0014 	&	1	&	0.2050 	&	$	8.6318 	^{+	0.0036 	}_{-	0.0037 	}	$	&	0.0556 	&	0.0589 	&	9.2611 	&	9.2356 	\\	
$	^{240}	$Am$	\to	^{236}	$Np$^{\rm{p}}$$	$	&	5.4700 	$\pm$	0.0500 	&	0	&	0.2083 	&	$	10.9834 	^{+	0.0026 	}_{-	0.0026 	}	$	&	0.0193 	&	0.0242 	&	12.1905 	&	12.0927 	\\	
$	^{241}	$Am$	\to	^{237}	$Np$	$	&	5.6378 	$\pm$	0.0001 	&	1	&	0.2116 	&	$	10.1352 	^{+	0.0006 	}_{-	0.0006 	}	$	&	0.0566 	&	0.0598 	&	10.7932 	&	10.7694 	\\	
$	^{242}	$Am$^{\rm{m}}$$	\to	^{238}	$Np$	$	&	5.6400 	$\pm$	0.0800 	&	3	&	0.2149 	&	$	11.9951 	^{+	0.0061 	}_{-	0.0062 	}	$	&	0.0072 	&	0.0070 	&	12.0641 	&	12.0747 	\\	
$	^{243}	$Am$	\to	^{239}	$Np$	$	&	5.4391 	$\pm$	0.0009 	&	1	&	0.2181 	&	$	11.3654 	^{+	0.0005 	}_{-	0.0005 	}	$	&	0.0568 	&	0.0587 	&	11.9377 	&	11.9236 	\\	
$	^{236}	$Cm$	\to	^{232}	$Pu$	$	&	7.0670 	$\pm$	0.0050 	&	0	&	0.1864 	&	$	3.3554 	^{+	0.0483 	}_{-	0.0544 	}	$	&	0.3230 	&	0.3648 	&	3.6574 	&	3.6045 	\\	
$	^{238}	$Cm$	\to	^{234}	$Pu$	$	&	6.6700 	$\pm$	0.0100 	&	0	&	0.1933 	&	$	5.3144 	^{+	0.0207 	}_{-	0.0217 	}	$	&	0.3306 	&	0.3718 	&	5.2900 	&	5.2389 	\\	
$	^{240}	$Cm$	\to	^{236}	$Pu$	$	&	6.3978 	$\pm$	0.0006 	&	0	&	0.2000 	&	$	6.4194 	^{+	0.0499 	}_{-	0.0564 	}	$	&	0.3329 	&	0.3654 	&	6.4949 	&	6.4545 	\\	
$	^{241}	$Cm$	\to	^{237}	$Pu$	$	&	6.1852 	$\pm$	0.0006 	&	3	&	0.2033 	&	$	8.4524 	^{+	0.0026 	}_{-	0.0027 	}	$	&	0.0304 	&	0.0311 	&	9.0389 	&	9.0294 	\\	
$	^{242}	$Cm$	\to	^{238}	$Pu$	$	&	6.2156 	$\pm$	0.0001 	&	0	&	0.2066 	&	$	7.1482 	^{+	0.0005 	}_{-	0.0005 	}	$	&	0.3307 	&	0.3484 	&	7.3414 	&	7.3188 	\\	
$	^{243}	$Cm$	\to	^{239}	$Pu$	$	&	6.1688 	$\pm$	0.0010 	&	2	&	0.2099 	&	$	8.9630 	^{+	0.0015 	}_{-	0.0015 	}	$	&	0.0382 	&	0.0372 	&	8.7588 	&	8.7707 	\\	
$	^{244}	$Cm$	\to	^{240}	$Pu$	$	&	5.9016 	$\pm$	0.0000 	&	0	&	0.2131 	&	$	8.7570 	^{+	0.0007 	}_{-	0.0007 	}	$	&	0.3378 	&	0.3562 	&	8.9451 	&	8.9221 	\\	
$	^{245}	$Cm$	\to	^{241}	$Pu$	$	&	5.6245 	$\pm$	0.0005 	&	2	&	0.2163 	&	$	11.4155 	^{+	0.0037 	}_{-	0.0037 	}	$	&	0.0410 	&	0.0429 	&	11.6296 	&	11.6103 	\\	
$	^{246}	$Cm$	\to	^{242}	$Pu$	$	&	5.4751 	$\pm$	0.0009 	&	0	&	0.2195 	&	$	11.1719 	^{+	0.0037 	}_{-	0.0037 	}	$	&	0.3555 	&	0.3927 	&	11.3118 	&	11.2686 	\\	
$	^{247}	$Cm$	\to	^{243}	$Pu$	$	&	5.3540 	$\pm$	0.0030 	&	1	&	0.2227 	&	$	14.6922 	^{+	0.0137 	}_{-	0.0141 	}	$	&	0.0550 	&	0.0597 	&	12.9422 	&	12.9063 	\\	
$	^{248}	$Cm$	\to	^{244}	$Pu$	$	&	5.1618 	$\pm$	0.0003 	&	0	&	0.2258 	&	$	13.0787 	^{+	0.0074 	}_{-	0.0076 	}	$	&	0.3680 	&	0.4166 	&	13.2399 	&	13.1860 	\\	
$	^{245}	$Bk$	\to	^{241}	$Am$	$	&	6.4545 	$\pm$	0.0014 	&	2	&	0.2082 	&	$	8.5519 	^{+	0.0026 	}_{-	0.0026 	}	$	&	0.0357 	&	0.0357 	&	7.8168 	&	7.8174 	\\	
$	^{247}	$Bk$	\to	^{243}	$Am$	$	&	5.8900 	$\pm$	0.0050 	&	2	&	0.2146 	&	$	10.6390 	^{+	0.0723 	}_{-	0.0868 	}	$	&	0.0382 	&	0.0409 	&	10.6341 	&	10.6049 	\\	
$	^{249}	$Bk$	\to	^{245}	$Am$	$	&	5.5210 	$\pm$	0.0014 	&	2	&	0.2209 	&	$	12.2900 	^{+	0.0004 	}_{-	0.0004 	}	$	&	0.0398 	&	0.0439 	&	12.7048 	&	12.6621 	\\	
$	^{238}	$Cf$	\to	^{234}	$Cm$	$	&	8.1300 	$\pm$	0.3000 	&	0	&	0.1765 	&	$	-0.0737 	^{+	0.0260 	}_{-	0.0276 	}	$	&	0.2796 	&	0.3417 	&	0.6053 	&	0.5182 	\\	
$	^{239}	$Cf$	\to	^{235}	$Cm$	$	&	7.7700 	$\pm$	0.0600 	&	0	&	0.1799 	&	$	2.7482 	^{+	0.0300 	}_{-	0.0322 	}	$	&	0.0639 	&	0.0819 	&	2.4864 	&	2.3784 	\\	
$	^{240}	$Cf$	\to	^{236}	$Cm$	$	&	7.7110 	$\pm$	0.0040 	&	0	&	0.1833 	&	$	1.6119 	^{+	0.0096 	}_{-	0.0098 	}	$	&	0.2839 	&	0.3404 	&	1.9924 	&	1.9134 	\\	
$	^{241}	$Cf$	\to	^{237}	$Cm$^{\rm{p}}$$	$	&	7.4600 	$\pm$	0.1500 	&	0	&	0.1867 	&	$	2.9731 	^{+	0.0321 	}_{-	0.0346 	}	$	&	0.0641 	&	0.0794 	&	3.5732 	&	3.4801 	\\	
$	^{242}	$Cf$	\to	^{238}	$Cm$	$	&	7.5170 	$\pm$	0.0040 	&	0	&	0.1901 	&	$	2.5356 	^{+	0.0183 	}_{-	0.0191 	}	$	&	0.2803 	&	0.3181 	&	2.6872 	&	2.6323 	\\	
$	^{243}	$Cf$	\to	^{239}	$Cm$	$	&	7.4200 	$\pm$	0.1000 	&	3	&	0.1934 	&	$	3.6654 	^{+	0.0119 	}_{-	0.0122 	}	$	&	0.0250 	&	0.0254 	&	4.5798 	&	4.5717 	\\	
$	^{244}	$Cf$	\to	^{240}	$Cm$	$	&	7.3290 	$\pm$	0.0018 	&	0	&	0.1967 	&	$	3.1931 	^{+	0.0110 	}_{-	0.0113 	}	$	&	0.2769 	&	0.2984 	&	3.5219 	&	3.4896 	\\	
$	^{245}	$Cf$	\to	^{241}	$Cm$	$	&	7.2585 	$\pm$	0.0018 	&	0	&	0.2000 	&	$	3.8836 	^{+	0.0142 	}_{-	0.0147 	}	$	&	0.0610 	&	0.0658 	&	4.2930 	&	4.2603 	\\	
$	^{246}	$Cf$	\to	^{242}	$Cm$	$	&	6.8616 	$\pm$	0.0010 	&	0	&	0.2033 	&	$	5.1090 	^{+	0.0060 	}_{-	0.0061 	}	$	&	0.2863 	&	0.3134 	&	5.3011 	&	5.2619 	\\	
$	^{247}	$Cf$	\to	^{243}	$Cm$	$	&	6.5030 	$\pm$	0.0140 	&	2	&	0.2065 	&	$	7.5050 	^{+	0.0042 	}_{-	0.0042 	}	$	&	0.0349 	&	0.0380 	&	8.0281 	&	7.9911 	\\	
$	^{248}	$Cf$	\to	^{244}	$Cm$	$	&	6.3610 	$\pm$	0.0050 	&	0	&	0.2097 	&	$	7.4596 	^{+	0.0349 	}_{-	0.0379 	}	$	&	0.3003 	&	0.3415 	&	7.5303 	&	7.4744 	\\	
$	^{249}	$Cf$	\to	^{245}	$Cm$	$	&	6.2933 	$\pm$	0.0005 	&	1	&	0.2129 	&	$	10.0444 	^{+	0.0025 	}_{-	0.0025 	}	$	&	0.0457 	&	0.0500 	&	8.7381 	&	8.6992 	\\	
$	^{250}	$Cf$	\to	^{246}	$Cm$	$	&	6.1285 	$\pm$	0.0002 	&	0	&	0.2160 	&	$	8.6160 	^{+	0.0030 	}_{-	0.0030 	}	$	&	0.3018 	&	0.3365 	&	8.7031 	&	8.6558 	\\	
$	^{251}	$Cf$	\to	^{247}	$Cm$	$	&	6.1770 	$\pm$	0.0009 	&	5	&	0.2191 	&	$	10.4524 	^{+	0.0208 	}_{-	0.0218 	}	$	&	0.0155 	&	0.0141 	&	10.8795 	&	10.9190 	\\	
$	^{252}	$Cf$	\to	^{248}	$Cm$	$	&	6.2170 	$\pm$	0.0000 	&	0	&	0.2222 	&	$	7.9352 	^{+	0.0013 	}_{-	0.0013 	}	$	&	0.2855 	&	0.2871 	&	8.1892 	&	8.1867 	\\	
$	^{253}	$Cf$	\to	^{249}	$Cm$^{\rm{m}}$$	$	&	6.1260 	$\pm$	0.0040 	&	0	&	0.2253 	&	$	9.0732 	^{+	0.0022 	}_{-	0.0022 	}	$	&	0.0634 	&	0.0647 	&	9.3318 	&	9.3223 	\\	
$	^{254}	$Cf$	\to	^{250}	$Cm$	$	&	5.9270 	$\pm$	0.0050 	&	0	&	0.2283 	&	$	9.2269 	^{+	0.0014 	}_{-	0.0014 	}	$	&	0.2908 	&	0.2937 	&	9.6619 	&	9.6576 	\\	
$	^{240}	$Es$	\to	^{236}	$Bk$	$	&	8.2600 	$\pm$	0.0600 	&	1	&	0.1750 	&	$	0.9331 	^{+	0.1083 	}_{-	0.1447 	}	$	&	0.0092 	&	0.0122 	&	2.0658 	&	1.9420 	\\	
$	^{241}	$Es$	\to	^{237}	$Bk$	$	&	8.2590 	$\pm$	0.0170 	&	0	&	0.1784 	&	$	0.7076 	^{+	0.0633 	}_{-	0.0741 	}	$	&	0.0588 	&	0.0757 	&	1.1736 	&	1.0642 	\\	
$	^{242}	$Es$	\to	^{238}	$Bk$	$	&	8.1600 	$\pm$	0.0200 	&	1	&	0.1818 	&	$	1.4945 	^{+	0.0374 	}_{-	0.0409 	}	$	&	0.0090 	&	0.0110 	&	2.3781 	&	2.2887 	\\	
$	^{243}	$Es$	\to	^{239}	$Bk$	$	&	8.0720 	$\pm$	0.0100 	&	0	&	0.1852 	&	$	1.5591 	^{+	0.0267 	}_{-	0.0284 	}	$	&	0.0579 	&	0.0701 	&	1.7670 	&	1.6843 	\\	
$	^{244}	$Es$	\to	^{240}	$Bk$	$	&	7.9400 	$\pm$	0.1000 	&	1	&	0.1885 	&	$	2.8692 	^{+	0.0446 	}_{-	0.0497 	}	$	&	0.0089 	&	0.0103 	&	3.1074 	&	3.0410 	\\	
$	^{245}	$Es$	\to	^{241}	$Bk$^{\rm{p}}$$	$	&	7.9090 	$\pm$	0.0030 	&	0	&	0.1918 	&	$	2.1333 	^{+	0.0229 	}_{-	0.0241 	}	$	&	0.0569 	&	0.0647 	&	2.2965 	&	2.2405 	\\	
$	^{246}	$Es$	\to	^{242}	$Bk$^{\rm{p}}$$	$	&	7.5000 	$\pm$	0.1000 	&	0	&	0.1951 	&	$	3.6576 	^{+	0.0280 	}_{-	0.0300 	}	$	&	0.0131 	&	0.0159 	&	4.4252 	&	4.3406 	\\	
$	^{246}	$Es$	\to	^{242}	$Bk$	$	&	7.6400 	$\pm$	0.1000 	&	1	&	0.1951 	&	$	3.6576 	^{+	0.0280 	}_{-	0.0300 	}	$	&	0.0089 	&	0.0100 	&	4.1407 	&	4.0883 	\\	
$	^{247}	$Es$	\to	^{243}	$Bk$^{\rm{p}}$$	$	&	7.4640 	$\pm$	0.0200 	&	0	&	0.1984 	&	$	3.5911 	^{+	0.0241 	}_{-	0.0256 	}	$	&	0.0582 	&	0.0662 	&	3.9122 	&	3.8565 	\\	
$	^{248}	$Es$	\to	^{244}	$Bk$	$	&	7.1600 	$\pm$	0.0500 	&	2	&	0.2016 	&	$	5.7604 	^{+	0.0512 	}_{-	0.0580 	}	$	&	0.0070 	&	0.0077 	&	6.2477 	&	6.2039 	\\	
$	^{249}	$Es$	\to	^{245}	$Bk$	$	&	6.9400 	$\pm$	0.0300 	&	3	&	0.2048 	&	$	6.0317 	^{+	0.0025 	}_{-	0.0026 	}	$	&	0.0244 	&	0.0256 	&	6.8656 	&	6.8460 	\\	
$	^{251}	$Es$	\to	^{247}	$Bk$	$	&	6.5971 	$\pm$	0.0010 	&	0	&	0.2112 	&	$	7.3758 	^{+	0.0130 	}_{-	0.0134 	}	$	&	0.0618 	&	0.0723 	&	7.5104 	&	7.4426 	\\	
$	^{252}	$Es$	\to	^{248}	$Bk$	$	&	6.7386 	$\pm$	0.0005 	&	2	&	0.2143 	&	$	7.7181 	^{+	0.0017 	}_{-	0.0018 	}	$	&	0.0069 	&	0.0071 	&	8.0043 	&	7.9905 	\\	
$	^{253}	$Es$	\to	^{249}	$Bk$	$	&	6.7392 	$\pm$	0.0001 	&	0	&	0.2174 	&	$	6.2476 	^{+	0.0006 	}_{-	0.0006 	}	$	&	0.0581 	&	0.0605 	&	6.8195 	&	6.8018 	\\	
$	^{254}	$Es$^{\rm{m}}$$	\to	^{250}	$Bk$	$	&	6.7000 	$\pm$	0.0011 	&	1	&	0.2205 	&	$	7.6455 	^{+	0.0022 	}_{-	0.0022 	}	$	&	0.0088 	&	0.0088 	&	7.8676 	&	7.8696 	\\	
$	^{255}	$Es$	\to	^{251}	$Bk$^{\rm{m}}$$	$	&	6.4363 	$\pm$	0.0013 	&	0	&	0.2235 	&	$	7.6333 	^{+	0.0129 	}_{-	0.0133 	}	$	&	0.0589 	&	0.0612 	&	8.2161 	&	8.1993 	\\	
$	^{241}	$Fm$	\to	^{237}	$Cf$	$	&	8.8600 	$\pm$	0.3200 	&	0	&	0.1701 	&	$	-2.2828 	^{+	0.0343 	}_{-	0.0372 	}	$	&	0.0559 	&	0.0786 	&	-0.2976 	&	-0.4457 	\\	
$	^{243}	$Fm$	\to	^{239}	$Cf$	$	&	8.6900 	$\pm$	0.0500 	&	1	&	0.1770 	&	$	-0.5954 	^{+	0.0166 	}_{-	0.0173 	}	$	&	0.0378 	&	0.0473 	&	0.4186 	&	0.3210 	\\	
$	^{244}	$Fm$	\to	^{240}	$Cf$	$	&	8.5500 	$\pm$	0.2000 	&	0	&	0.1803 	&	$	-0.5058 	^{+	0.0110 	}_{-	0.0113 	}	$	&	0.2457 	&	0.3049 	&	-0.0521 	&	-0.1460 	\\	
$	^{245}	$Fm$	\to	^{241}	$Cf$	$	&	8.4400 	$\pm$	0.1000 	&	3	&	0.1837 	&	$	0.6232 	^{+	0.1171 	}_{-	0.1609 	}	$	&	0.0219 	&	0.0243 	&	1.7756 	&	1.7295 	\\	
$	^{246}	$Fm$	\to	^{242}	$Cf$	$	&	8.3790 	$\pm$	0.0050 	&	0	&	0.1870 	&	$	0.2181 	^{+	0.0111 	}_{-	0.0114 	}	$	&	0.2412 	&	0.2809 	&	0.4677 	&	0.4016 	\\	
$	^{247}	$Fm$^{\rm{m}}$$	\to	^{243}	$Cf$	$	&	8.3100 	$\pm$	0.0080 	&	0	&	0.1903 	&	$	0.7631 	^{+	0.0167 	}_{-	0.0174 	}	$	&	0.0530 	&	0.0616 	&	1.3301 	&	1.2653 	\\	
$	^{247}	$Fm$	\to	^{243}	$Cf$	$	&	8.2580 	$\pm$	0.0100 	&	4	&	0.1903 	&	$	1.6852 	^{+	0.0138 	}_{-	0.0142 	}	$	&	0.0165 	&	0.0167 	&	2.7547 	&	2.7485 	\\	
$	^{248}	$Fm$	\to	^{244}	$Cf$	$	&	7.9950 	$\pm$	0.0080 	&	0	&	0.1935 	&	$	1.5378 	^{+	0.0148 	}_{-	0.0154 	}	$	&	0.2436 	&	0.2781 	&	1.7242 	&	1.6668 	\\	
$	^{250}	$Fm$	\to	^{246}	$Cf$	$	&	7.5570 	$\pm$	0.0080 	&	0	&	0.2000 	&	$	3.2695 	^{+	0.0151 	}_{-	0.0157 	}	$	&	0.2490 	&	0.2839 	&	3.2958 	&	3.2388 	\\	
$	^{251}	$Fm$	\to	^{247}	$Cf$	$	&	7.4255 	$\pm$	0.0010 	&	1	&	0.2032 	&	$	6.0253 	^{+	0.0065 	}_{-	0.0066 	}	$	&	0.0382 	&	0.0422 	&	4.6670 	&	4.6234 	\\	
$	^{252}	$Fm$	\to	^{248}	$Cf$	$	&	7.1537 	$\pm$	0.0010 	&	0	&	0.2063 	&	$	4.9610 	^{+	0.0007 	}_{-	0.0007 	}	$	&	0.2545 	&	0.2908 	&	4.8532 	&	4.7953 	\\	
$	^{253}	$Fm$	\to	^{249}	$Cf$^{\rm{m}}$$	$	&	7.1979 	$\pm$	0.0010 	&	5	&	0.2095 	&	$	6.3345 	^{+	0.0170 	}_{-	0.0177 	}	$	&	0.0131 	&	0.0123 	&	7.0770 	&	7.1049 	\\	
$	^{254}	$Fm$	\to	^{250}	$Cf$	$	&	7.3073 	$\pm$	0.0010 	&	0	&	0.2126 	&	$	4.0671 	^{+	0.0003 	}_{-	0.0003 	}	$	&	0.2392 	&	0.2436 	&	4.2029 	&	4.1951 	\\	
$	^{256}	$Fm$	\to	^{252}	$Cf$	$	&	7.0253 	$\pm$	0.0019 	&	0	&	0.2188 	&	$	5.0658 	^{+	0.0036 	}_{-	0.0036 	}	$	&	0.2406 	&	0.2411 	&	5.3409 	&	5.3401 	\\	
$	^{257}	$Fm$	\to	^{253}	$Cf$	$	&	6.8637 	$\pm$	0.0009 	&	2	&	0.2218 	&	$	6.9396 	^{+	0.0009 	}_{-	0.0009 	}	$	&	0.0283 	&	0.0270 	&	7.2221 	&	7.2430 	\\	
$	^{244}	$Md$	\to	^{240}	$Es$	$	&	8.9500 	$\pm$	0.0800 	&	1	&	0.1721 	&	$	-0.4437 	^{+	0.1427 	}_{-	0.2139 	}	$	&	0.0082 	&	0.0116 	&	0.6558 	&	0.5029 	\\	
$	^{245}	$Md$	\to	^{241}	$Es$	$	&	9.0100 	$\pm$	0.1200 	&	2	&	0.1755 	&	$	-0.4202 	^{+	0.1015 	}_{-	0.1326 	}	$	&	0.0272 	&	0.0343 	&	0.0963 	&	-0.0039 	\\	
$	^{246}	$Md$	\to	^{242}	$Es$	$	&	8.8900 	$\pm$	0.0400 	&	1	&	0.1789 	&	$	-0.0362 	^{+	0.0776 	}_{-	0.0946 	}	$	&	0.0079 	&	0.0103 	&	0.8100 	&	0.6947 	\\	
$	^{246}	$Md$^{\rm{m}}$$	\to	^{242}	$Es$	$	&	8.9500 	$\pm$	0.0600 	&	3	&	0.1789 	&	$	0.8876 	^{+	0.0726 	}_{-	0.0872 	}	$	&	0.0046 	&	0.0055 	&	1.2445 	&	1.1621 	\\	
$	^{247}	$Md$^{\rm{m}}$$	\to	^{243}	$Es$	$	&	9.0300 	$\pm$	0.0400 	&	3	&	0.1822 	&	$	-0.4997 	^{+	0.0645 	}_{-	0.0757 	}	$	&	0.0200 	&	0.0221 	&	0.3371 	&	0.2936 	\\	
$	^{247}	$Md$	\to	^{243}	$Es$	$	&	8.7640 	$\pm$	0.0100 	&	1	&	0.1822 	&	$	0.0755 	^{+	0.0317 	}_{-	0.0342 	}	$	&	0.0354 	&	0.0436 	&	0.5689 	&	0.4780 	\\	
$	^{249}	$Md$^{\rm{m}}$$	\to	^{245}	$Es$	$	&	8.5400 	$\pm$	0.1000 	&	2	&	0.1888 	&	$	0.2788 	^{+	0.1684 	}_{-	0.2788 	}	$	&	0.0266 	&	0.0301 	&	1.4521 	&	1.3989 	\\	
$	^{249}	$Md$	\to	^{245}	$Es$	$	&	8.4410 	$\pm$	0.0180 	&	2	&	0.1888 	&	$	1.5332 	^{+	0.0150 	}_{-	0.0155 	}	$	&	0.0269 	&	0.0310 	&	1.7682 	&	1.7075 	\\	
$	^{250}	$Md$	\to	^{246}	$Es$	$	&	8.1550 	$\pm$	0.2800 	&	2	&	0.1920 	&	$	2.8873 	^{+	0.0310 	}_{-	0.0334 	}	$	&	0.0061 	&	0.0073 	&	3.3631 	&	3.2864 	\\	
$	^{251}	$Md$	\to	^{247}	$Es$	$	&	7.9630 	$\pm$	0.0040 	&	1	&	0.1952 	&	$	3.4024 	^{+	0.0231 	}_{-	0.0244 	}	$	&	0.0362 	&	0.0430 	&	3.0905 	&	3.0157 	\\	
$	^{253}	$Md$	\to	^{249}	$Es$	$	&	7.5730 	$\pm$	0.0080 	&	1	&	0.2016 	&	$	5.0122 	^{+	0.2218 	}_{-	0.4771 	}	$	&	0.0368 	&	0.0433 	&	4.4898 	&	4.4183 	\\	
$	^{255}	$Md$	\to	^{251}	$Es$	$	&	7.9056 	$\pm$	0.0017 	&	2	&	0.2078 	&	$	4.3644 	^{+	0.0310 	}_{-	0.0334 	}	$	&	0.0257 	&	0.0253 	&	3.5082 	&	3.5159 	\\	
$	^{256}	$Md$	\to	^{252}	$Es$	$	&	7.7400 	$\pm$	0.1100 	&	3	&	0.2109 	&	$	4.7048 	^{+	0.0099 	}_{-	0.0102 	}	$	&	0.0044 	&	0.0043 	&	5.1077 	&	5.1200 	\\	
$	^{257}	$Md$	\to	^{253}	$Es$	$	&	7.5571 	$\pm$	0.0009 	&	1	&	0.2140 	&	$	5.1222 	^{+	0.0039 	}_{-	0.0040 	}	$	&	0.0341 	&	0.0343 	&	4.5126 	&	4.5103 	\\	
$	^{258}	$Md$	\to	^{254}	$Es$	$	&	7.2713 	$\pm$	0.0019 	&	1	&	0.2171 	&	$	6.6491 	^{+	0.0024 	}_{-	0.0024 	}	$	&	0.0078 	&	0.0082 	&	6.2798 	&	6.2550 	\\	
$	^{258}	$Md$^{\rm{m}}$$	\to	^{254}	$Es$^{\rm{m}}$$	$	&	7.1900 	$\pm$	0.2000 	&	1	&	0.2171 	&	$	5.4548 	^{+	0.0068 	}_{-	0.0069 	}	$	&	0.0079 	&	0.0085 	&	6.6269 	&	6.5943 	\\	
$	^{251}	$No$	\to	^{247}	$Fm$	$	&	8.7520 	$\pm$	0.0040 	&	0	&	0.1873 	&	$	-0.0160 	^{+	0.0054 	}_{-	0.0055 	}	$	&	0.0485 	&	0.0627 	&	0.6727 	&	0.5611 	\\	
$	^{251}	$No$^{\rm{m}}$$	\to	^{247}	$Fm$^{\rm{m}}$$	$	&	8.8200 	$\pm$	0.0060 	&	0	&	0.1873 	&	$	0.0086 	^{+	0.0126 	}_{-	0.0130 	}	$	&	0.0481 	&	0.0615 	&	0.4596 	&	0.3530 	\\	
$	^{252}	$No$	\to	^{248}	$Fm$	$	&	8.5490 	$\pm$	0.0050 	&	0	&	0.1905 	&	$	0.5622 	^{+	0.0028 	}_{-	0.0028 	}	$	&	0.2191 	&	0.2723 	&	0.6253 	&	0.5309 	\\	
$	^{253}	$No$	\to	^{249}	$Fm$	$	&	8.4150 	$\pm$	0.0040 	&	1	&	0.1937 	&	$	2.2337 	^{+	0.0055 	}_{-	0.0056 	}	$	&	0.0335 	&	0.0402 	&	1.9772 	&	1.8983 	\\	
$	^{254}	$No$	\to	^{250}	$Fm$	$	&	8.2260 	$\pm$	0.0080 	&	0	&	0.1969 	&	$	1.7550 	^{+	0.0034 	}_{-	0.0034 	}	$	&	0.2196 	&	0.2652 	&	1.6615 	&	1.5796 	\\	
$	^{255}	$No$	\to	^{251}	$Fm$^{\rm{m}}$$	$	&	8.2320 	$\pm$	0.0030 	&	2	&	0.2000 	&	$	2.8476 	^{+	0.0217 	}_{-	0.0228 	}	$	&	0.0252 	&	0.0277 	&	2.7869 	&	2.7451 	\\	
$	^{256}	$No$	\to	^{252}	$Fm$	$	&	8.5820 	$\pm$	0.0050 	&	0	&	0.2031 	&	$	0.4663 	^{+	0.0074 	}_{-	0.0075 	}	$	&	0.2021 	&	0.2107 	&	0.4826 	&	0.4646 	\\	
$	^{257}	$No$	\to	^{253}	$Fm$	$	&	8.4770 	$\pm$	0.0060 	&	2	&	0.2062 	&	$	1.4597 	^{+	0.0088 	}_{-	0.0090 	}	$	&	0.0235 	&	0.0228 	&	1.9517 	&	1.9647 	\\	
$	^{259}	$No$	\to	^{255}	$Fm$	$	&	7.8540 	$\pm$	0.0050 	&	2	&	0.2124 	&	$	3.6665 	^{+	0.0359 	}_{-	0.0392 	}	$	&	0.0245 	&	0.0247 	&	4.0591 	&	4.0563 	\\	
$	^{252}	$Lr$	\to	^{248}	$Md$	$	&	9.1640 	$\pm$	0.0170 	&	0	&	0.1825 	&	$	-0.4242 	^{+	0.0804 	}_{-	0.0987 	}	$	&	0.0103 	&	0.0147 	&	0.4522 	&	0.2978 	\\	
$	^{253}	$Lr$	\to	^{249}	$Md$	$	&	8.9180 	$\pm$	0.3000 	&	0	&	0.1858 	&	$	-0.1535 	^{+	0.0306 	}_{-	0.0329 	}	$	&	0.0467 	&	0.0643 	&	0.5053 	&	0.3662 	\\	
$	^{253}	$Lr$^{\rm{m}}$$	\to	^{249}	$Md$^{\rm{m}}$$	$	&	8.8600 	$\pm$	0.1000 	&	0	&	0.1858 	&	$	0.1663 	^{+	0.0438 	}_{-	0.0487 	}	$	&	0.0470 	&	0.0653 	&	0.6851 	&	0.5419 	\\	
$	^{254}	$Lr$	\to	^{250}	$Md$	$	&	8.8200 	$\pm$	0.0080 	&	3	&	0.1890 	&	$	1.2237 	^{+	0.0314 	}_{-	0.0339 	}	$	&	0.0041 	&	0.0051 	&	2.2813 	&	2.1899 	\\	
$	^{255}	$Lr$	\to	^{251}	$Md$^{\rm{p}}$$	$	&	8.5560 	$\pm$	0.0070 	&	0	&	0.1922 	&	$	1.4941 	^{+	0.0151 	}_{-	0.0156 	}	$	&	0.0469 	&	0.0629 	&	1.6150 	&	1.4874 	\\	
$	^{255}	$Lr$^{\rm{m}}$$	\to	^{251}	$Md$	$	&	8.6000 	$\pm$	0.0800 	&	0	&	0.1922 	&	$	0.8028 	^{+	0.0085 	}_{-	0.0086 	}	$	&	0.0467 	&	0.0621 	&	1.4662 	&	1.3419 	\\	
$	^{255}	$Lr$^{\rm{p}}$$	\to	^{251}	$Md$	$	&	10.0200 	$\pm$	0.0220 	&	9	&	0.1922 	&	$	0.0743 	^{+	0.0120 	}_{-	0.0124 	}	$	&	0.0033 	&	0.0026 	&	1.6752 	&	1.7793 	\\	
$	^{256}	$Lr$	\to	^{252}	$Md$	$	&	8.8500 	$\pm$	0.1200 	&	1	&	0.1953 	&	$	1.5162 	^{+	0.0153 	}_{-	0.0159 	}	$	&	0.0068 	&	0.0082 	&	1.5709 	&	1.4904 	\\	
$	^{257}	$Lr$	\to	^{253}	$Md$^{\rm{p}}$$	$	&	9.0700 	$\pm$	0.0300 	&	4	&	0.1984 	&	$	0.7782 	^{+	0.0280 	}_{-	0.0300 	}	$	&	0.0131 	&	0.0129 	&	1.2370 	&	1.2463 	\\	
$	^{259}	$Lr$	\to	^{255}	$Md$^{\rm{p}}$$	$	&	8.5800 	$\pm$	0.0700 	&	0	&	0.2046 	&	$	0.9003 	^{+	0.0021 	}_{-	0.0021 	}	$	&	0.0433 	&	0.0490 	&	1.4812 	&	1.4278 	\\	
$	^{255}	$Rf$	\to	^{251}	$No$	$	&	9.0550 	$\pm$	0.0040 	&	1	&	0.1843 	&	$	0.4896 	^{+	0.0131 	}_{-	0.0135 	}	$	&	0.0311 	&	0.0438 	&	0.6844 	&	0.5358 	\\	
$	^{256}	$Rf$	\to	^{252}	$No$	$	&	8.9260 	$\pm$	0.0150 	&	0	&	0.1875 	&	$	0.3282 	^{+	0.0033 	}_{-	0.0033 	}	$	&	0.2020 	&	0.2824 	&	0.1746 	&	0.0292 	\\	
$	^{257}	$Rf$^{\rm{m}}$$	\to	^{253}	$No$	$	&	9.1600 	$\pm$	0.0110 	&	2	&	0.1907 	&	$	0.7063 	^{+	0.0189 	}_{-	0.0197 	}	$	&	0.0226 	&	0.0277 	&	0.6343 	&	0.5455 	\\	
$	^{257}	$Rf$	\to	^{253}	$No$^{\rm{m}}$$	$	&	9.0830 	$\pm$	0.0080 	&	2	&	0.1907 	&	$	0.7481 	^{+	0.0170 	}_{-	0.0177 	}	$	&	0.0228 	&	0.0283 	&	0.8361 	&	0.7420 	\\	
$	^{258}	$Rf$	\to	^{254}	$No$	$	&	9.1960 	$\pm$	0.0130 	&	0	&	0.1938 	&	$	-0.5933 	^{+	0.0170 	}_{-	0.0177 	}	$	&	0.1886 	&	0.2315 	&	-0.6432 	&	-0.7322 	\\	
$	^{259}	$Rf$	\to	^{255}	$No$	$	&	9.1300 	$\pm$	0.0700 	&	2	&	0.1969 	&	$	0.4905 	^{+	0.0409 	}_{-	0.0452 	}	$	&	0.0218 	&	0.0247 	&	0.7037 	&	0.6496 	\\	
$	^{261}	$Rf$	\to	^{257}	$No$	$	&	8.6500 	$\pm$	0.0700 	&	0	&	0.2031 	&	$	1.0669 	^{+	0.0395 	}_{-	0.0435 	}	$	&	0.0422 	&	0.0517 	&	1.6194 	&	1.5317 	\\	
$	^{261}	$Rf$^{\rm{m}}$$	\to	^{257}	$No$^{\rm{p}}$$	$	&	8.4200 	$\pm$	0.1000 	&	1	&	0.2031 	&	$	1.8692 	^{+	0.0498 	}_{-	0.0563 	}	$	&	0.0300 	&	0.0365 	&	2.6501 	&	2.5644 	\\	
$	^{256}	$Db$	\to	^{252}	$Lr$	$	&	9.3400 	$\pm$	0.0300 	&	2	&	0.1797 	&	$	0.3854 	^{+	0.0918 	}_{-	0.1165 	}	$	&	0.0051 	&	0.0078 	&	1.1239 	&	0.9384 	\\	
$	^{257}	$Db$	\to	^{253}	$Lr$	$	&	9.2060 	$\pm$	0.0200 	&	1	&	0.1829 	&	$	0.3886 	^{+	0.0362 	}_{-	0.0395 	}	$	&	0.0300 	&	0.0451 	&	0.5794 	&	0.4026 	\\	
$	^{257}	$Db$^{\rm{m}}$$	\to	^{253}	$Lr$^{\rm{m}}$$	$	&	9.3100 	$\pm$	0.1100 	&	0	&	0.1829 	&	$	-0.1134 	^{+	0.0372 	}_{-	0.0407 	}	$	&	0.0430 	&	0.0666 	&	0.0564 	&	-0.1330 	\\	
$	^{258}	$Db$	\to	^{254}	$Lr$	$	&	9.4370 	$\pm$	0.1100 	&	1	&	0.1860 	&	$	0.5303 	^{+	0.0772 	}_{-	0.0648 	}	$	&	0.0064 	&	0.0091 	&	0.5766 	&	0.4217 	\\	
$	^{259}	$Db$	\to	^{255}	$Lr$^{\rm{m}}$$	$	&	9.6200 	$\pm$	0.0500 	&	1	&	0.1892 	&	$	-0.2924 	^{+	0.1185 	}_{-	0.1635 	}	$	&	0.0276 	&	0.0357 	&	-0.5885 	&	-0.6995 	\\	
$	^{263}	$Db$	\to	^{259}	$Lr$	$	&	8.8300 	$\pm$	0.1500 	&	5	&	0.2015 	&	$	1.8942 	^{+	0.1174 	}_{-	0.1614 	}	$	&	0.0096 	&	0.0105 	&	3.1240 	&	3.0871 	\\	
$	^{259}	$Sg$^{\rm{m}}$$	\to	^{255}	$Rf$^{\rm{m}}$$	$	&	9.7100 	$\pm$	0.0220 	&	2	&	0.1815 	&	$	-0.6327 	^{+	0.0490 	}_{-	0.0553 	}	$	&	0.0213 	&	0.0312 	&	-0.1723 	&	-0.3391 	\\	
$	^{259}	$Sg$	\to	^{255}	$Rf$	$	&	9.7650 	$\pm$	0.0080 	&	2	&	0.1815 	&	$	-0.3958 	^{+	0.0566 	}_{-	0.0651 	}	$	&	0.0212 	&	0.0308 	&	-0.3763 	&	-0.5395 	\\	
$	^{260}	$Sg$	\to	^{256}	$Rf$	$	&	9.9010 	$\pm$	0.0100 	&	0	&	0.1846 	&	$	-1.7678 	^{+	0.0280 	}_{-	0.0300 	}	$	&	0.1751 	&	0.2513 	&	-1.9063 	&	-2.0630 	\\	
$	^{261}	$Sg$	\to	^{257}	$Rf$	$	&	9.7140 	$\pm$	0.0150 	&	2	&	0.1877 	&	$	-0.7292 	^{+	0.0117 	}_{-	0.0120 	}	$	&	0.0205 	&	0.0275 	&	-0.2630 	&	-0.3917 	\\	
$	^{263}	$Sg$	\to	^{259}	$Rf$	$	&	9.4000 	$\pm$	0.0600 	&	0	&	0.1939 	&	$	0.0336 	^{+	0.0603 	}_{-	0.0700 	}	$	&	0.0388 	&	0.0545 	&	0.0897 	&	-0.0577 	\\	
$	^{263}	$Sg$^{\rm{m}}$$	\to	^{259}	$Rf$^{\rm{m}}$$	$	&	9.4600 	$\pm$	0.0190 	&	2	&	0.1939 	&	$	-0.3768 	^{+	0.0928 	}_{-	0.1181 	}	$	&	0.0203 	&	0.0261 	&	0.4046 	&	0.2956 	\\	
$	^{265}	$Sg$	\to	^{261}	$Rf$	$	&	9.0500 	$\pm$	0.1200 	&	5	&	0.2000 	&	$	1.2648 	^{+	0.0696 	}_{-	0.0830 	}	$	&	0.0092 	&	0.0106 	&	2.7859 	&	2.7261 	\\	
$	^{261}	$Bh$	\to	^{257}	$Db$	$	&	10.5000 	$\pm$	0.0700 	&	3	&	0.1801 	&	$	-1.8928 	^{+	0.0969 	}_{-	0.1249 	}	$	&	0.0148 	&	0.0205 	&	-1.6293 	&	-1.7710 	\\	
$	^{263}	$Hs$^{\rm{m}}$$	\to	^{261}	$Sg$	$	&	10.4700 	$\pm$	0.1100 	&	0	&	0.1787 	&	$	-3.0000 	^{+	0.0000 	}_{-	0.0000 	}	$	&	0.0361 	&	0.0612 	&	-2.0303 	&	-2.2595 	\\	
$	^{263}	$Hs$	\to	^{259}	$Sg$^{\rm{m}}$$	$	&	10.7300 	$\pm$	0.0800 	&	2	&	0.1787 	&	$	-3.0458 	^{+	0.1597 	}_{-	0.2553 	}	$	&	0.0185 	&	0.0280 	&	-2.1807 	&	-2.3605 	\\	
$	^{265}	$Hs$	\to	^{259}	$Sg$^{\rm{m}}$$	$	&	10.4700 	$\pm$	0.0800 	&	0	&	0.1849 	&	$	-2.7077 	^{+	0.0341 	}_{-	0.0370 	}	$	&	0.0348 	&	0.0541 	&	-2.0546 	&	-2.2463 	\\	
$	^{266}	$Hs$	\to	^{262}	$Sg$	$	&	10.3460 	$\pm$	0.0160 	&	0	&	0.1880 	&	$	-2.4037 	^{+	0.0792 	}_{-	0.0969 	}	$	&	0.1555 	&	0.2288 	&	-2.4251 	&	-2.5929 	\\	
$	^{268}	$Hs$	\to	^{264}	$Sg$	$	&	9.7600 	$\pm$	0.1000 	&	0	&	0.1940 	&	$	0.1461 	^{+	0.2518 	}_{-	0.6690 	}	$	&	0.1590 	&	0.2346 	&	-0.9024 	&	-1.0712 	\\	
$	^{269}	$Hs$	\to	^{265}	$Sg$	$	&	9.2700 	$\pm$	0.1700 	&	1	&	0.1970 	&	$	1.1761 	^{+	0.1663 	}_{-	0.2730 	}	$	&	0.0253 	&	0.0381 	&	1.3660 	&	1.1881 	\\	
$	^{270}	$Hs$	\to	^{266}	$Sg$	$	&	9.0700 	$\pm$	0.0400 	&	0	&	0.2000 	&	$	0.9542 	^{+	0.1597 	}_{-	0.2553 	}	$	&	0.1656 	&	0.2518 	&	1.0836 	&	0.9017 	\\	
$	^{267}	$Ds$	\to	^{263}	$Hs$	$	&	11.7800 	$\pm$	0.0500 	&	0	&	0.1760 	&	$	-5.0000 	^{+	0.2553 	}_{-	0.6990 	}	$	&	0.0310 	&	0.0527 	&	-4.3858 	&	-4.6168 	\\	
$	^{270}	$Ds$	\to	^{266}	$Hs$	$	&	11.1170 	$\pm$	0.0280 	&	0	&	0.1852 	&	$	-3.6882 	^{+	0.0914 	}_{-	0.1159 	}	$	&	0.1395 	&	0.2197 	&	-3.6451 	&	-3.8425 	\\	
$	^{282}	$Ds$	\to	^{278}	$Hs$	$	&	9.1500 	$\pm$	0.4200 	&	0	&	0.2199 	&	$	2.4014 	^{+	0.2518 	}_{-	0.6690 	}	$	&	0.1373 	&	0.1830 	&	1.4340 	&	1.3093 	\\	
$	^{286}	$Cn$	\to	^{282}	$Ds$	$	&	9.2400 	$\pm$	0.7600 	&	0	&	0.2168 	&	$	1.4771 	^{+	0.3010 	}_{-	0.3010 	}	$	&	0.1317 	&	0.2058 	&	1.8639 	&	1.6699 	\\	
$	^{286}	$Fl$	\to	^{282}	$Cn$	$	&	10.3600 	$\pm$	0.0400 	&	0	&	0.2028 	&	$	-0.6569 	^{+	0.0099 	}_{-	0.0101 	}	$	&	0.1209 	&	0.2165 	&	-0.6521 	&	-0.9052 	\\	
$	^{288}	$Fl$	\to	^{284}	$Cn$	$	&	10.0760 	$\pm$	0.0120 	&	0	&	0.2083 	&	$	-0.1851 	^{+	0.0693 	}_{-	0.0825 	}	$	&	0.1202 	&	0.2091 	&	0.1033 	&	-0.1370 	\\	
$	^{290}	$Lv$	\to	^{286}	$Fl$	$	&	11.0000 	$\pm$	0.0600 	&	0	&	0.2000 	&	$	-2.0458 	^{+	0.1249 	}_{-	0.1761 	}	$	&	0.1097 	&	0.2121 	&	-1.6569 	&	-1.9432 	\\	
$	^{292}	$Lv$	\to	^{288}	$Fl$	$	&	10.7910 	$\pm$	0.0120 	&	0	&	0.2055 	&	$	-1.7959 	^{+	0.1383 	}_{-	0.2041 	}	$	&	0.1081 	&	0.2003 	&	-1.1615 	&	-1.4293 	\\	
$	^{294}	$Og$	\to	^{290}	$Lv$	$	&	11.8700 	$\pm$	0.0300 	&	0	&	0.1973 	&	$	-3.1549 	^{+	0.1549 	}_{-	0.2430 	}	$	&	0.0980 	&	0.1992 	&	-3.0551 	&	-3.3633 	\\	
\toprule
\label{table1}
\end{longtable*}

\begin{figure*}[!htbp]\centering
 \includegraphics
  [width=0.8\hsize]
  {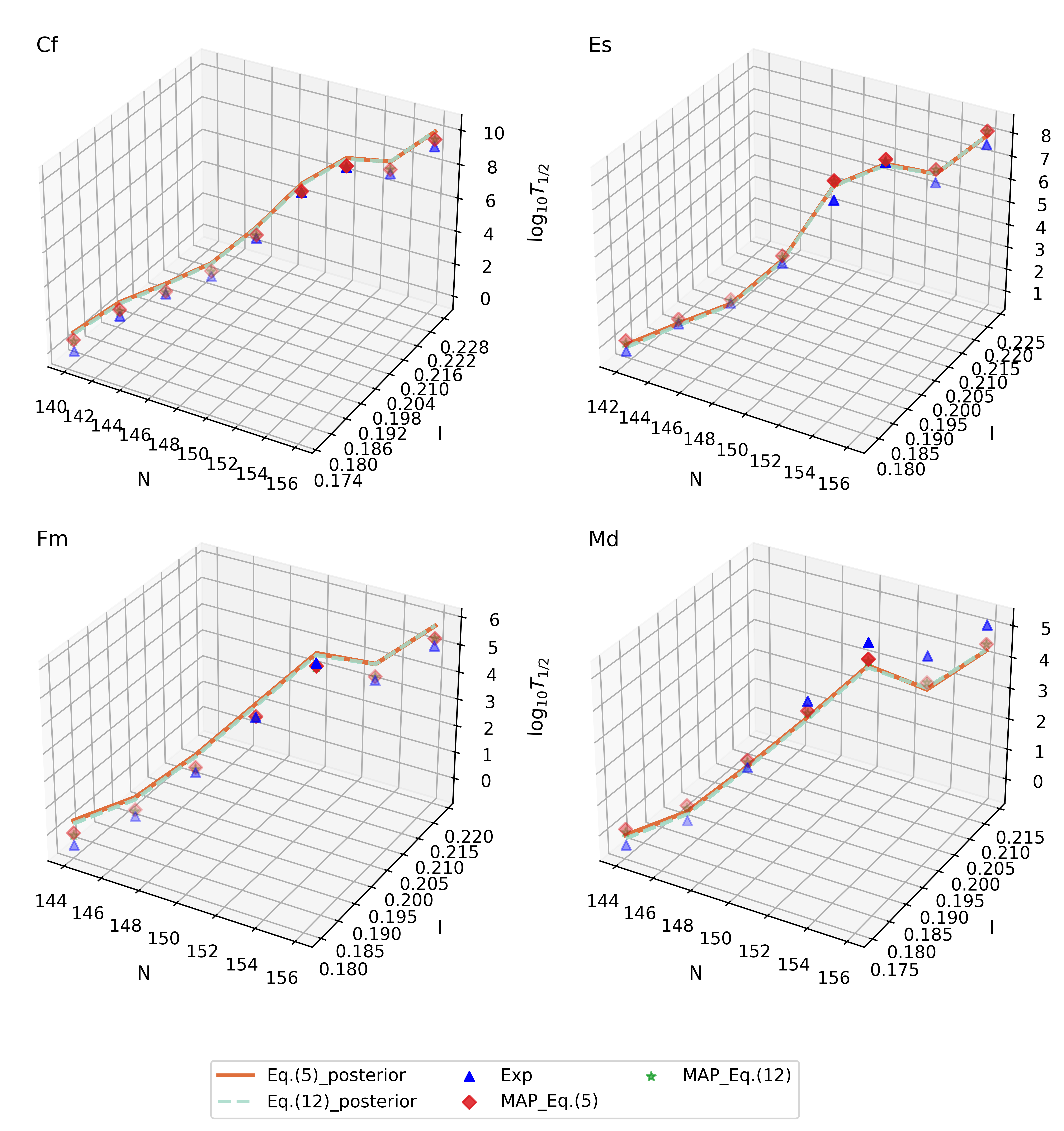}
 \caption{The $\alpha$ decay half-lives of Cf, Es, Fm, and Md near $N=152$, as calculated using the posterior distributions of Eq. (\ref{Eq5}) and Eq. (\ref{Eq12}) along with their corresponding MAP values.}
 \label{Fig5}
 \end{figure*}
 
 \begin{figure*}[!htbp]\centering
 \includegraphics
  [width=0.8\hsize]
  {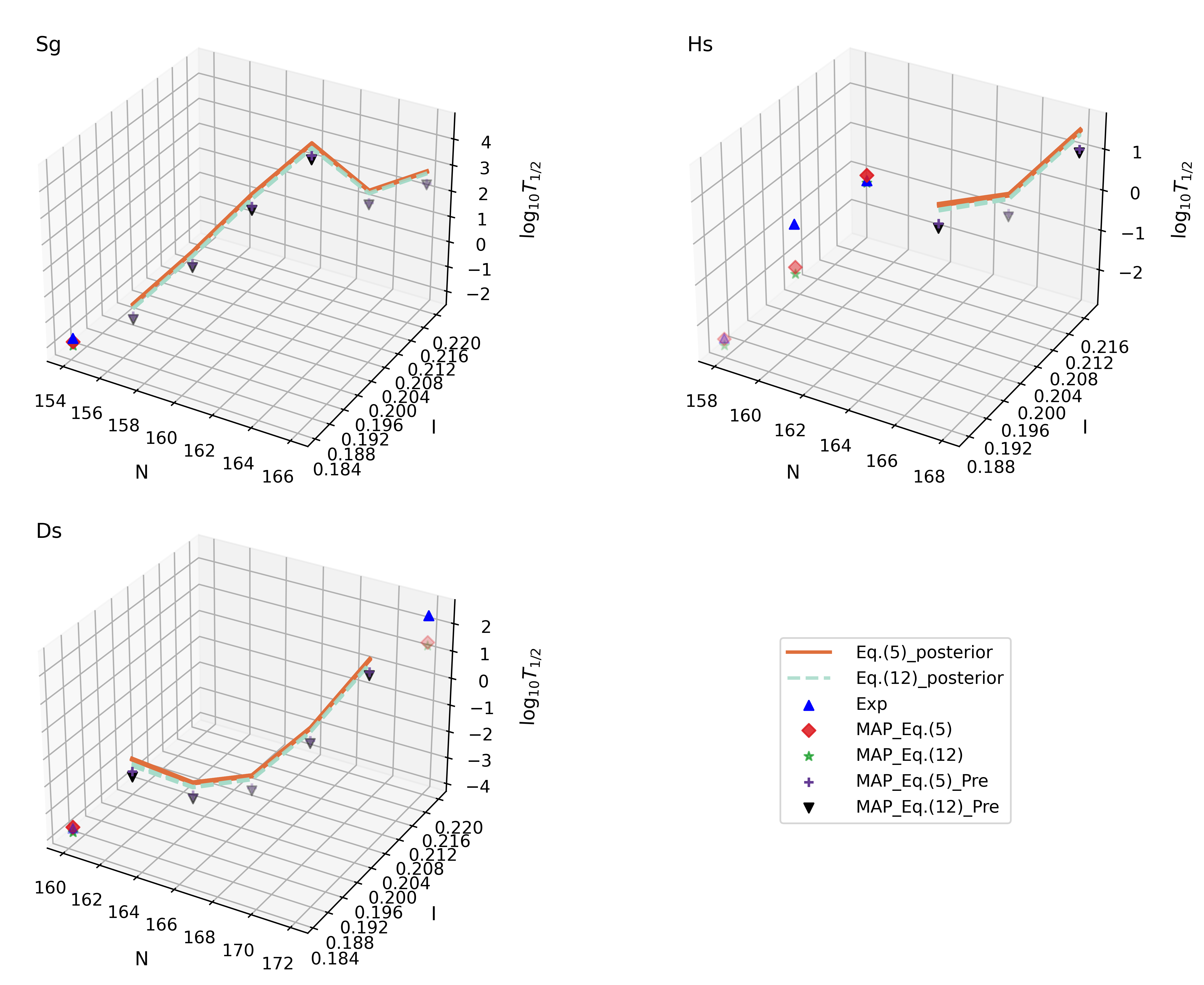}
 \caption{The predicted half-lives of Sg, Hs, and Ds near $N=162$, as calculated using the posterior distributions of Eq. (\ref{Eq5}) and Eq. (\ref{Eq12}) along with their corresponding MAP values.}
 \label{Fig6}
 \end{figure*}

Leveraging the robustness of posterior distributions, we further constructed the three-dimensional plot shown in Fig. \ref{Fig5}, based on the isospin $I$, neutron number $N$, and the logarithm of the $\alpha$ decay half-life, $\mathrm{log_{10}}T_{1/2}$. The figure comprises four subplots, corresponding to the four superheavy nuclides Cf, Es, Fm, and Md. Each subplot uses a white grid as the background, with $N$ on the horizontal axis, $I$ on the depth axis, and $\mathrm{log_{10}}T_{1/2}$ on the vertical axis. Plotted within are the posterior distribution curves derived from Eq. (\ref{Eq5}) and Eq. (\ref{Eq12}). Experimental ones are denoted by blue triangles, while the red diamonds and green asterisks represent the calculations based on the MAP values from Eq. (\ref{Eq5}) and Eq. (\ref{Eq12}), respectively. The difference between the posterior curve and the MAP estimate value stems from the fact that MAP is a single point estimate. Due to the asymmetric shape of the posterior distribution, especially in regions with high uncertainty such as near the closure area of the shell layer, it may not be consistent with the posterior mean or median.
A clear observation from the figure is a pronounced shell effect near $N=152$ for all nuclides, where the half-life curves exhibit distinct peaks. Furthermore, the predictions from the posterior distributions show good overall agreement with the experimental ones, thereby providing further validation of the Bayesian calibration capability to describe the $\alpha$ decay behavior of nuclides with different parities.
Meanwhile, Table \ref{table2} presents predictions of the half-lives and preformation factors for superheavy nuclei based on the $\alpha$ decay enetgy $Q_{\alpha}$ from the WS4 mass model. As can be seen from Table \ref{table2}, for $^{262}\rm{Sg}$, the $P_{\alpha}$ predicted by Eq. (\ref{Eq5}) is 0.1731, while that predicted by Eq. (\ref{Eq12}) with considering the isospin effect is 0.2367, with corresponding $\rm{log_{10}}T_{1/2}$  of -1.2510 and -1.3868, respectively. This discrepancy highlights the influence of isospin correction on the preformation factor. However, around $N=162$, the predicted values exhibit a relatively wide variation, indicating our model's sensitivity to input parameters. To more clearly visualize the shell effects near $N=162$ in superheavy nuclei, Fig. \ref{Fig6} displays the predicted  $\alpha$ decay half-lives for Sg, Hs, and Ds isotopes around $N=162$ based on Bayesian posterior distributions. This figure shows that despite scarce experimental data, the posterior distributions from Eq. (\ref{Eq5}) and Eq. (\ref{Eq12}) are still able to capture the overall trend of half-lives. Nevertheless, near $N=162$ a broad confidence interval appears, which precisely reflects the structural complexity of superheavy nuclei, for instance, nuclear deformation may blur shell effects, while collective motion may obscure single-particle effects. Through prior constraints and global fitting, the Bayesian approach can partially capture the trend of shell effects. However, its reliance on $Q_{\alpha}$ inputs from the WS4 mass model increases the uncertainty of parameter extrapolation, particularly in regions with limited data.

In addition, Table \ref{table3} lists the preformation factors and $\alpha$ decay half-lives for both medium-heavy and heavy nuclei with different parities. These values are calculated using Eq. (\ref{Eq5}) and, with the inclusion of the isospin effect, Eq. (\ref{Eq12}), and are subsequently compared with experimental ones as well as results from the Refs. \cite{Wan:2021wny} and \cite{Deng:2021siq}. The table clearly shows that the $\alpha$ decay half-lives computed with the present model, particularly those obtained from the corrected isospin effect Eq. (\ref{Eq12}), are in remarkable agreement with the experimental values and remain consistent with other theoretical findings. This comparative analysis further confirms the effectiveness of the current model in characterizing $\alpha$ decay behavior across various types of atomic nuclei.

\begin{table*}[!htbp]
    \centering
    \setlength\tabcolsep{2.8pt}
    \renewcommand\arraystretch{0.85}
    \caption{The predicted $\alpha$-particle preformation factors and half-lives of Sg, Hs, and Ds near $N$ = 162 are calculated using Eq. (\ref{Eq5})and Eq. (\ref{Eq12}) with  the $Q_{\alpha}$ values extracted from the WS4 mass model \cite{Zhu:2025ujz}.}
  \begin{tabular}{ccccccc}	\hline
      \toprule
       $\alpha$ transition \, \,\,\,\,&$\rm{Q_{\alpha}^{WS4}}$\, \,\,&I\, \,\,\,\,       &$\rm{P_{\alpha}^{{Cal1}}}$\, \,\,&$\rm{P_{\alpha}^{{Cal2}}}$
         \, \,\,\,\,&${\rm{log}}_{10}T_{1/2}^{\rm{{Cal1}}}$ \, \,\,\,\, &${\rm{log}}_{10}T_{1/2}^{\rm{{Cal2}}}$\, \,\,\,\,	                 
         \\\hline
$	^{262}	$Sg$	\to	^{258}	$Rf$	$	&	9.65 	&	\,\,\,	0.1908 	&	\,\,\,	0.1731 	&	\,\,\,	0.2367 	&	-1.2510 	&	-1.3868 	\\
$	^{264}	$Sg$	\to	^{260}	$Rf$	$	&	9.05 	&	\,\,\,	0.1970 	&	\,\,\,	0.1783 	&	\,\,\,	0.2466 	&	0.4662 	&	0.3252 	\\
$	^{266}	$Sg$	\to	^{262}	$Rf$	$	&	8.44 	&	\,\,\,	0.2030 	&	\,\,\,	0.1851 	&	\,\,\,	0.2630 	&	2.3801 	&	2.2276 	\\
$	^{268}	$Sg$	\to	^{264}	$Rf$	$	&	7.98 	&	\,\,\,	0.2090 	&	\,\,\,	0.1898 	&	\,\,\,	0.2726 	&	4.0214 	&	3.8642 	\\
$	^{270}	$Sg$	\to	^{266}	$Rf$	$	&	8.56 	&	\,\,\,	0.2148 	&	\,\,\,	0.1693 	&	\,\,\,	0.2020 	&	1.9398 	&	1.8631 	\\
$	^{272}	$Sg$	\to	^{268}	$Rf$	$	&	8.42 	&	\,\,\,	0.2206 	&	\,\,\,	0.1662 	&	\,\,\,	0.1891 	&	2.4229 	&	2.3669 	\\
$	^{272}	$Hs$	\to	^{268}	$Sg$	$	&	9.53 	&	\,\,\,	0.2059 	&	\,\,\,	0.1515 	&	\,\,\,	0.1979 	&	-0.3085 	&	-0.4245 	\\
$	^{274}	$Hs$	\to	^{270}	$Sg$	$	&	9.50 	&	\,\,\,	0.2117 	&	\,\,\,	0.1465 	&	\,\,\,	0.1784 	&	-0.2801 	&	-0.3656 	\\
$	^{276}	$Hs$	\to	^{272}	$Sg$	$	&	9.05 	&	\,\,\,	0.2174 	&	\,\,\,	0.1488 	&	\,\,\,	0.1813 	&	1.0742 	&	0.9884 	\\
$	^{272}	$Ds$	\to	^{268}	$Hs$	$	&	10.38 	&	\,\,\,	0.1912 	&	\,\,\,	0.1440 	&	\,\,\,	0.2306 	&	-1.8923 	&	-2.0968 	\\
$	^{274}	$Ds$	\to	^{270}	$Hs$	$	&	10.87 	&	\,\,\,	0.1971 	&	\,\,\,	0.1326 	&	\,\,\,	0.1835 	&	-3.1044 	&	-3.2456 	\\
$	^{276}	$Ds$	\to	^{272}	$Hs$	$	&	10.88 	&	\,\,\,	0.2029 	&	\,\,\,	0.1277 	&	\,\,\,	0.1635 	&	-3.1514 	&	-3.2588 	\\
$	^{278}	$Ds$	\to	^{274}	$Hs$	$	&	10.25 	&	\,\,\,	0.2086 	&	\,\,\,	0.1309 	&	\,\,\,	0.1695 	&	-1.6569 	&	-1.7692 	\\
$	^{280}	$Ds$	\to	^{276}	$Hs$	$	&	9.43 	&	\,\,\,	0.2143 	&	\,\,\,	0.1378 	&	\,\,\,	0.1881 	&	0.6111 	&	0.4761 	\\
\toprule
\end{tabular}
   \label{table2}
\end{table*}

 \begin{table*}[!htbp]
    \centering
    \setlength\tabcolsep{1.8pt}
    \renewcommand\arraystretch{0.85}
    \caption{The $\alpha$-particle preformation factors $P_{\alpha}^{\rm{Cal1}}$ and $P_{\alpha}^{\rm{Cal2}}$, are extracted from Eq. (\ref{Eq5}) and the included isospin effect Eq. (\ref{Eq12}), respectively. Comparing with both the experimental half-lives $\mathrm{log_{10}}T_{1/2}^{\rm{exp}}$ and the calculated half-lives $\mathrm{log_{10}}T_{1/2}^{\rm{Cal1}}$, $\mathrm{log_{10}}T_{1/2}^{\rm{Cal2}}$, $\mathrm{log_{10}}T_{1/2}^{\rm{Cal3}}$, and $\mathrm{log_{10}}T_{1/2}^{\rm{Cal4}}$ obtained using Eq. (\ref{Eq5}), Eq. (\ref{Eq12}), and Refs. \cite{Wan:2021wny} and \cite{Deng:2021siq}.}
  \begin{tabular}{cccccccccc}	\hline
        Nuclei &$l$ &\,$\rm{Q_{\alpha}}$\, \,&$P_{\alpha}^{\rm{Cal1}}$\, \,&$P_{\alpha}^{\rm{Cal2}}$
         \, \,&$\mathrm{log_{10}}T_{1/2}^{\rm{exp}}$ &\, \,$\mathrm{log_{10}}T_{1/2}^{\rm{Cal1}}$ \, \, &$\mathrm{log_{10}}T_{1/2}^{\rm{Cal2}}$\,\,	                  &\, \,$\mathrm{log_{10}}T_{1/2}^{\rm{Cal3}}$&\, \,$\mathrm{log_{10}}T_{1/2}^{\rm{Cal4}}$\\
                        &    &   &  \, \,  Eq. (\ref{Eq5}) &  \, \,  Eq. (\ref{Eq12}) &   & \, \,  Eq. (\ref{Eq5})     &\, \, Eq. (\ref{Eq12})     &\, \,   \cite{Wan:2021wny}  & \, \, \cite{Deng:2021siq} \\\hline
$^{105}_{52}$Te$$&2 &\,\,5.069 	&\, \,0.502 &\, \,1.198 &-6.1986 &-6.7258 &-6.1037&-6.6203&-\\
$^{109}_{53}$I$$&2 &\,\,3.918 	&\,\,0.546 	&\,\,0.803 	&-0.1786&-1.2748&-1.4427 &-1.1491&-\\
$^{210}_{82}$Pb$$&0 &\,\,3.792 	&\,\,0.796 	&\,\,0.507 	&16.5667&15.6790&15.8750&- & 15.18\\
$^{212}_{84}$Po$$ &	0 &\,\,8.954 &\,\,0.329 &\,\,0.081 	&-6.5311&-6.7680&-6.1605&-&-6.87\\\hline
\end{tabular}
   \label{table3}
\end{table*}

\section{Summary} \label{Sec.IV}

In our systematic investigation of the correlation between isospin asymmetry and $\alpha$-particle preformation probability in the superheavy region, a Bayesian inference method is employed. Within the CTP framework, $P_{\alpha}$ are first extracted from experimental decay energies and half-lives, leading to the construction of a multi-parameter model incorporating an isospin asymmetry term. To efficiently explore the high-dimensional parameter space and perform uncertainty quantification, a GP emulator is constructed using design points generated via LHS. Posterior distributions of the parameters are subsequently obtained through MCMC sampling based on the Metropolis-Hastings algorithm. It is shown that neutron-proton asymmetry significantly suppresses $P_{\alpha}$.
Furthermore, an independent analysis using the random forest method confirm the prominent role of isospin, showing that it exhibits a higher relative dependence among the various influencing factors. These findings collectively demonstrate that the Bayesian-calibrated model is capable of providing a comprehensive description of $\alpha$ decay preformation factors and half-lives for superheavy nuclei with different structural types, within the framework established by the CTP-based data. 
This work offers a self-consistent theoretical method for systematically exploring decay properties in the superheavy region and underscores the necessity of incorporating isospin asymmetry corrections in $\alpha$ decay studies.


\section*{References}
\begin{thebibliography} {99}

\bibitem{Gamow:1928zz} G. Gamow, Zur Quantentheorie des Atomkernes. Z. Phys. {\bf 51}, 204--212 (1928).   \href{https://doi.org/10.1007/BF01343196}{https://doi.org/10.1007/BF01343196}

\bibitem{Gurney:1928lxa} R.W. Gurney, E.~U. Condon, Wave Mechanics and Radioactive Disintegration. Nature {\bf 122}, 439--439 (1928).  
    \href{https://doi.org/10.1038/122439a0}{https://doi.org/10.1038/122439a0}

\bibitem{Mang:1964yi} H.J. Mang, Alpha decay, Ann. Rev. Nucl. Part. Sci. {\bf 14} 1--26 (1964).
\href {https://doi.org/10.1146/annurev.ns.14.120164.000245}
{https://doi.org/10.1146/annurev.ns.14.120164.000245}

\bibitem{Andreyev:2013iwa} A.N. Andreyev, M. Huyse, P. Van Duppen et~al., Signatures of the Z=82 Shell Closure in
  {\ensuremath{\alpha}}-Decay Process, Phys. Rev. Lett. {\bf110} 242502 (2013).
\href {https://doi.org/10.1103/PhysRevLett.110.242502}
  {https://doi.org/10.1103/PhysRevLett.110.242502}

\bibitem{Hamilton:2013una} J.H. Hamilton, S. Hofmann, Y.T. Oganessian, Search for Superheavy Nuclei,
  Ann. Rev. Nucl. Part. Sci. {\bf63} 383--405 (2013).
\href {https://doi.org/10.1146/annurev-nucl-102912-144535}
  {https://doi.org/10.1146/annurev-nucl-102912-144535}
  
\bibitem{Singh:1992hbk} S. Singh, R.K. Gupta, W. Scheid, et~al., {Possible synthesis of new and
  superheavy elements via cluster decay}, J. Phys. G {\bf18} 1243--1257 (1992).
 \href {https://doi.org/10.1088/0954-3899/18/7/012}
  {{doi:10.1088/0954-3899/18/7/012}}.

\bibitem{Oganessian:2007zza} Y.T. Oganessian, V.K. Utyonkov, Yu.V. Lobanov et~al., {Synthesis of the isotope \ce{^{282}113} in the \ce{^{237}Np}+\ce{^{48}Ca} fusion reaction}, Phys. Rev. C {\bf76} 011601 (2007).
\href {https://doi.org/10.1103/PhysRevC.76.011601}
  {https://doi.org/10.1103/PhysRevC.76.011601}

\bibitem{Oganessian:2010zz} Y.T. Oganessian, F.Sh. Abdullin, P.D. Bailey et~al., {Synthesis of a New Element with Atomic Number
  Z=117}, Phys. Rev. Lett. {\bf104} 142502 (2010).
\href {https://doi.org/10.1103/PhysRevLett.104.142502}
  {https://doi.org/10.1103/PhysRevLett.104.142502}

\bibitem{Ellison:2010zz} P.A. Ellison, K.E. Gregorich, J.S. Berryman et~al., {New Superheavy Element Isotopes: \ce{^{242}Pu} (\ce{^{48}Ca}, 5n)\ce{^{285}114}
  128514}, Phys. Rev. Lett. {\bf105} 182701 (2010).
\href {https://doi.org/10.1103/PhysRevLett.105.182701}
  {https://doi.org/10.1103/PhysRevLett.105.182701}

\bibitem{ALICE:2012dtf} B. Abelev, C. Alice, {Upgrade of the ALICE Experiment: Letter Of Intent}, J.
  Phys. G {\bf41} 087001 (2014).
\href {https://doi.org/10.1088/0954-3899/41/8/087001}
  {https://doi.org/10.1088/0954-3899/41/8/087001}

\bibitem{Ma:2015qga} L. Ma, Z.Y. Zhang, Z.G. Gan et~al., {\ensuremath{\alpha}-decay properties of the new isotope
 \ce{^{216}U}}, Phys. Rev. C {\bf91} 051302 (2015).
\href {https://doi.org/10.1103/PhysRevC.91.051302}
  {https://doi.org/10.1103/PhysRevC.91.051302}

\bibitem{Hofmann:2000cs} S. Hofmann, G. Munzenberg, {The Discovery of the Heaviest Elements}, Rev. Mod. Phys. {\bf72} 733--767 (2000).
\href {https://doi.org/10.1103/RevModPhys.72.733}
  {https://doi.org/10.1103/RevModPhys.72.733}
  
\bibitem{Wang:2025} Z.X. Wang, G.X. Zhang et~al., {Lifetime of first excited state in $\ce{^{139}La}$ and the role of core-excitation on L-forbidden M1 transition}, Nucl. Sci. Tech. {\bf36} 179 (2025).
\href{https://doi.org/10.1007/s41365-025-01756-7}
{https://doi.org/10.1007/s41365-025-01756-7}  

\bibitem{Jia2026} Z.H. Jia, Y.D. Fang, S.C. Wang et~al., {HALIMA: a hybrid array for lifetime measurement of neutron-rich nuclei at IMP}, Nucl. Sci. Tech. {\bf37} 25 (2026).
\href{https://doi.org/10.1007/s41365-025-01830-0}{https://doi.org/10.1007/s41365-025-01830-0}

\bibitem{Dodig-Crnkovic:1989fpw} G. Dodig-Crnkovic, F.A. Janouch, R.J. Liotta, {An exact shell-model treatment of {\ensuremath{\alpha}}-clustering and absolute {\ensuremath{\alpha}}-decay}, Nucl. Phys. A {\bf501} 533--545 (1989).
\href {https://doi.org/10.1016/0375-9474(89)90146-2}
   {https://doi.org/10.1016/0375-9474(89)90146-2}

\bibitem{Varga:1992zz} K. Varga, R.G. Lovas, R.J. Liotta, {Absolute alpha decay width of \ce{^{212}Po} in a
  combined shell and cluster model}, Phys. Rev. Lett. {\bf69} 37--40 (1992).
\href {https://doi.org/10.1103/PhysRevLett.69.37}
  {https://doi.org/10.1103/PhysRevLett.69.37}

\bibitem{Royer:1985mtk} G. Royer, B. Remaud, {Static and dynamic fusion barriers in heavy-ion
  reactions}, Nucl. Phys. A {\bf444} 477--497 (1985).
\href {https://doi.org/10.1016/0375-9474(85)90464-6}
  {https://doi.org/10.1016/0375-9474(85)90464-6}

\bibitem{Zhang:2006dj} H. Zhang, W. Zuo, J. Li et al., {alpha decay half-lives of new superheavy
  nuclei within a generalized liquid drop model}, Phys. Rev. C {\bf74} 017304 (2006).
\href {https://doi.org/10.1103/PhysRevC.74.017304}
  {https://doi.org/10.1103/PhysRevC.74.017304}

\bibitem{Ropke:2014wsa} G. R{\"o}pke, P. Schuck, B. Zhou et al., {Nuclear clusters bound to doubly magic nuclei: The case of $^{212}$Po}, Phys. Rev. C {\bf90} 034304 (2014).
  \href {https://doi.org/10.1103/PhysRevC.90.034304}
   {https://doi.org/10.1103/PhysRevC.90.034304}

\bibitem{Xu:2015pvv} C. Xu, Z. Ren, G. R{\"o}pke et al., {$\alpha$-decay width of $^{212}$Po from a quartetting wave function approach}, Phys. Rev. C {\bf93} 011306 (2016).
  \href {https://doi.org/10.1103/PhysRevC.93.011306}
  {https://doi.org/10.1103/PhysRevC.93.011306}

\bibitem{buck1992favoured}
B.Buck, A.C. Merchant, S.M. Perez, {Favoured alpha decays of odd-mass
  nuclei}, J. Phys. G {\bf18} 143 (1992).
\href {https://doi.org/10.1088/0954-3899/18/1/012}
  {https://doi.org/10.1088/0954-3899/18/1/012}

\bibitem{Buck:1992zz}
B. Buck, A.C. Merchant, S.M. Perez, {alpha decay calculations with a
  realistic potential}, Phys. Rev. C {\bf45} 2247--2253 (1992).
\href {https://doi.org/10.1103/PhysRevC.45.2247}
  {https://doi.org/10.1103/PhysRevC.45.2247}

\bibitem{Poenaru:2016rbm} D.N. Poenaru, R.A. Gherghescu, {Spontaneous fission of the superheavy nucleus $^{286}$Fl}, Phys. Rev. C {\bf94} 014309 (2016).
  \href {https://doi.org/10.1103/PhysRevC.94.014309}
 {https://doi.org/10.1103/PhysRevC.94.014309}

\bibitem{Denisov:2005ax} V.Y. Denisov, H. Ikezoe, {Alpha-nucleus potential for alpha-decay and
  sub-barrier fusion}, Phys. Rev. C {\bf72} 064613 (2005).
\href {https://doi.org/10.1103/PhysRevC.72.064613}
  {https://doi.org/10.1103/PhysRevC.72.064613}

\bibitem{Chowdhury:2005nd} P.R. Chowdhury, C. Samanta, D.N. Basu, {Alpha decay half-lives of new
  superheavy elements}, Phys. Rev. C {\bf73} 014612 (2006).
\href{https://doi.org/10.1103/PhysRevC.73.014612}
  {https://doi.org/10.1103/PhysRevC.73.014612}

\bibitem{RoyChowdhury:2008uh} P. Roy~Chowdhury, C. Samanta, D.N. Basu, {Search for long lived heaviest
  nuclei beyond the valley of stability}, Phys. Rev. C {\bf77} 044603 (2008).
  \href {https://doi.org/10.1103/PhysRevC.77.044603}
  {https://doi.org/10.1103/PhysRevC.77.044603}

\bibitem{Poenaru:2018jju} D.N. Poenaru, H. St{\"o}cker, R.A. Gherghescu, {Cluster and alpha decay of
  superheavy nuclei}, Eur. Phys. J. A {\bf54} 14 (2018).
\href {https://doi.org/10.1140/epja/i2018-12469-6}
  {https://doi.org/10.1140/epja/i2018-12469-6}

\bibitem{Zhu:2024swx} X.Y. Zhu, S. Luo, W.~Gao et al., {An improved simple model for {\ensuremath{\alpha}} decay half-lives}, Chin. Phys. C {\bf48} 074102 (2024).
\href {https://doi.org/10.1088/1674-1137/ad3d4b}
  {https://doi.org/10.1088/1674-1137/ad3d4b}

\bibitem{Zhu:2025ujz} X.Y. Zhu, W. Gao, L. Zhu et al., {Bayesian inference correlation: From {\ensuremath{\alpha}} -particle preformation factor to {\ensuremath{\alpha}} -decay properties in heavy and superheavy nuclei}, Phys. Rev. C {\bf112} 024329 (2025).
\href {https://doi.org/10.1103/x53t-7fb8}
  {https://doi.org/10.1103/x53t-7fb8}
  
\bibitem{Gangopadhyay:2009dke} G.~Gangopadhyay, {Simple parametrization of $\alpha$-decay spectroscopic factor
  in $150\le A\le200$ region}, J. Phys. G {\bf36} 095105 (2009).
  \href {https://doi.org/10.1088/0954-3899/36/9/095105}
  {https://doi.org/10.1088/0954-3899/36/9/095105}

\bibitem{Guo:2014era} S.~Guo, X.~Bao, Y.~Gao, J.~Li et al., {The nuclear deformation and the
  preformation factor in the {\ensuremath{\alpha}}-decay of heavy and
  superheavy nuclei}, Nucl. Phys. A {\bf934} 110--120 (2014).
\href {https://doi.org/10.1016/j.nuclphysa.2014.12.001}
  {https://doi.org/10.1016/j.nuclphysa.2014.12.001}

\bibitem{Deng:2020rzy}J.G. Deng, H.F. Zhang, {Analytic formula for estimating the
  {\ensuremath{\alpha}}-particle preformation factor}, Phys. Rev. C {\bf102} 044314 (2020).
\href {https://doi.org/10.1103/PhysRevC.102.044314}
  {https://doi.org/10.1103/PhysRevC.102.044314}

\bibitem{Deng:2017ids} J.G. Deng, J.C. Zhao, D. Xiang et al., {Systematic study of unfavored
  {\ensuremath{\alpha}}-decay half-lives of closed-shell nuclei related to
  ground and isomeric states}, Phys. Rev. C {\bf96} 024318 (2017).
  \href {https://doi.org/10.1103/PhysRevC.96.024318}
 {https://doi.org/10.1103/PhysRevC.96.024318}

\bibitem{Deng:2018eva} J.G. Deng, J.C. Zhao, P.C. Chu et al., {Systematic study of
  {\ensuremath{\alpha}} decay of nuclei around $\boldsymbol{Z=82}$,
  $\boldsymbol{N=126}$ shell closure within the cluster-formation model and
  proximity potential 1977 formalism}, Phys. Rev. C {\bf97} 044322 (2018).
  \href {https://doi.org/10.1103/PhysRevC.97.044322}
  {https://doi.org/10.1103/PhysRevC.97.044322}

\bibitem{Deng:2021siq} J.G. Deng, H.F. Zhang, {Correlation between {\ensuremath{\alpha}}-particle
  preformation factor and {\ensuremath{\alpha}} decay energy}, Phys. Lett. B {\bf816} 136247 (2021).
\href {https://doi.org/10.1016/j.physletb.2021.136247}
  {https://doi.org/10.1016/j.physletb.2021.136247}

\bibitem{Seif:2011zz} W.M. Seif, M. Shalaby, M.F. Alrakshy, {Isospin asymmetry dependence of the alpha spectroscopic factor for heavy nuclei}, Phys. Rev. C {\bf84} 064608 (2011).
\href {https://doi.org/10.1103/PhysRevC.84.064608}
  {https://doi.org/10.1103/PhysRevC.84.064608}

\bibitem{Ma2025} Y.G Ma, {Multi-proton emission at the limits of nuclear stability: challenges for extreme open quantum systems}, Nucl. Sci. Tech. {\bf36} 236 (2025).
\href{https://doi.org/10.1007/s41365-025-01831-z}{https://doi.org/10.1007/s41365-025-01831-z}

\bibitem{Luo:2024ogt} S.~Luo, D.M. Zhang, L.J. Qi et al.,
  {{\ensuremath{\alpha}}-particle preformation factors in heavy and superheavy
  nuclei*}, Chin. Phys. C {\bf48} 044105 (2024).
\href {https://doi.org/10.1088/1674-1137/ad21e9}
  {https://doi.org/10.1088/1674-1137/ad21e9}

\bibitem{Huang:2021nwk} W.J. Huang, M.~Wang, F.G. Kondev et al., {The AME 2020 atomic
  mass evaluation (I). Evaluation of input data, and adjustment procedures},
  Chin. Phys. C {\bf45} 030002 (2021).
 \href {https://doi.org/10.1088/1674-1137/abddb0}
  {https://doi.org/10.1088/1674-1137/abddb0}

\bibitem{Wang:2021xhn} M.~Wang, W.J. Huang, F.G. Kondev et al., {The AME 2020 atomic
  mass evaluation (II). Tables, graphs and references}, Chin. Phys. C {\bf45} 030003
  (2021).
\href {https://doi.org/10.1088/1674-1137/abddaf}
   {https://doi.org/10.1088/1674-1137/abddaf}
  
\bibitem{tang1993orthogonal} B.~Tang, {Orthogonal array-based Latin hypercubes}, J. Am. Stat. Assoc.
  {\bf88} 1392--1397 (1993).
\href {https://doi.org/10.1080/01621459.1993.10476423}
 {https://doi.org/10.1080/01621459.1993.10476423}

\bibitem{morris1995exploratory} M.D. Morris, T.J. Mitchell, {Exploratory designs for computational
  experiments}, J. Stat. Plann. Infer. {\bf43} 381--402 (1995).
\href {https://doi.org/10.1016/0378-3758(94)00035-T}
  {https://doi.org/10.1016/0378-3758(94)00035-T}

\bibitem{Foreman-Mackey:2012any} D.~Foreman-Mackey, D.W. Hogg, D.~Lang et al., {emcee: The MCMC Hammer},
  Publ. Astron. Soc. Pac. {\bf125} 306--312 (2013).
  \href {https://doi.org/10.1086/670067} {https://doi.org/10.1086/670067}

\bibitem{mcmillan1999analysis} N.J. McMillan, J.~Sacks, W.~J. Welch et al., {Analysis of protein activity
  data by Gaussian stochastic process models}, J. Biopharm. Stat. {\bf9} 145--160 (1999).
\href {https://doi.org/10.1081/BIP-100101005}
  {https://doi.org/10.1081/BIP-100101005}
  
\bibitem{Jin:2025} S.L.Jin, J.G. Li, Y. Gao, et al., {Full configuration interaction quantum Monte Carlo in nuclear structure calculations.}, Nucl. Sci. Tech. {\bf36} 212 (2025). 
\href {https://doi.org/10.1007/s41365-025-01790-5}
  {https://doi.org/10.1007/s41365-025-01790-5}
  
\bibitem{He:2018gks} Y.~He, L.G. Pang, X.N. Wang, {Bayesian extraction of jet energy loss
  distributions in heavy-ion collisions}, Phys. Rev. Lett. {\bf122} 252302 (2019).
  \href {https://doi.org/10.1103/PhysRevLett.122.252302}
  {https://doi.org/10.1103/PhysRevLett.122.252302}

\bibitem{Wu:2023azi} J.~Wu, W.~Ke, X.N. Wang, {Bayesian inference of the path-length dependence of
  jet energy loss}, Phys. Rev. C {\bf108} 034911 (2023).
  \href {https://doi.org/10.1103/PhysRevC.108.034911}
  {https://doi.org/10.1103/PhysRevC.108.034911}

\bibitem{Xing:2023ciw} W.J. Xing, S.~Cao, G.Y. Qin, {Flavor hierarchy of parton energy loss in
  quark-gluon plasma from a Bayesian analysis}, Phys. Lett. B {\bf850} 138523 (2024).
  \href {https://doi.org/10.1016/j.physletb.2024.138523}
  {https://doi.org/10.1016/j.physletb.2024.138523}

\bibitem{OmanaKuttan:2022aml} M.~Omana~Kuttan, J.~Steinheimer, K.~Zhou et al., {QCD Equation of State
  of Dense Nuclear Matter from a Bayesian Analysis of Heavy-Ion Collision
  Data}, Phys. Rev. Lett. {\bf131} 202303 (2023).
  \href {https://doi.org/10.1103/PhysRevLett.131.202303}
  {https://doi.org/10.1103/PhysRevLett.131.202303}

\bibitem{Zhu:2025gxo} L.~Zhu, X.~Chen, K.~Zhou et al., {Bayesian inference of the
  critical end point in a (2+1)-flavor system from holographic QCD}, Phys. Rev.
  D {\bf112} 026019 (2025).
  \href {https://doi.org/10.1103/wpts-lbtr} {https://doi.org/10.1103/wpts-lbtr}

\bibitem{Cheng:2023ucp} Y.L. Cheng, S.~Shi, Y.G. Ma et al., {Examination of nucleon
  distribution with Bayesian imaging for isobar collisions}, Phys. Rev. C
  {\bf107} 064909 (2023).
\href {https://doi.org/10.1103/PhysRevC.107.064909}
  {https://doi.org/10.1103/PhysRevC.107.064909}

\bibitem{Novak:2013bqa}
J.~Novak, K.~Novak, S.~Pratt et al., {Determining Fundamental Properties of Matter Created in Ultrarelativistic Heavy-Ion Collisions}, Phys. Rev. C {\bf89} 034917 (2014).
\href {https://doi.org/10.1103/PhysRevC.89.034917}
  {https://doi.org/10.1103/PhysRevC.89.034917}

\bibitem{Pratt:2015zsa} S.~Pratt, E.~Sangaline, P.~Sorensen et al., {Constraining the Eq. of State of
  Super-Hadronic Matter from Heavy-Ion Collisions}, Phys. Rev. Lett. {\bf114} 202301 (2015).
\href {https://doi.org/10.1103/PhysRevLett.114.202301}
  {https://doi.org/10.1103/PhysRevLett.114.202301}

\bibitem{Sangaline:2015isa} E.~Sangaline, S.~Pratt, {Toward a deeper understanding of how experiments
  constrain the underlying physics of heavy-ion collisions}, Phys. Rev. C
  {\bf93} 024908 (2016).
\href {https://doi.org/10.1103/PhysRevC.93.024908}
  {https://doi.org/10.1103/PhysRevC.93.024908}

\bibitem{JETSCAPE:2020shq} D.~Everett, W. Ke, J.F. Paquetet et al., {Phenomenological constraints on the transport properties
  of QCD matter with data-driven model averaging}, Phys. Rev. Lett. {\bf126} 242301 (2021).
\href {https://doi.org/10.1103/PhysRevLett.126.242301}
  {https://doi.org/10.1103/PhysRevLett.126.242301}

\bibitem{Mantysaari:2022ffw} H.~M{\"a}ntysaari, B.~Schenke, C.~Shen et al., {Bayesian inference of the
  fluctuating proton shape}, Phys. Lett. B {\bf833} 137348 (2022).
\href {https://doi.org/10.1016/j.physletb.2022.137348}
  {https://doi.org/10.1016/j.physletb.2022.137348}

\bibitem{Goodman:2010dyf} J.~Goodman, J.~Weare, {Ensemble samplers with affine invariance}, Commun. Appl. Math. Comput. Sc. {\bf5} 65--80 (2010).
\href {https://doi.org/10.2140/camcos.2010.5.65}
  {https://doi.org/10.2140/camcos.2010.5.65}

\bibitem{Shin:2015iiw} E.~Shin, Y.~Lim, C.H. Hyun et al., {Nuclear isospin asymmetry in $\alpha$
  decay of heavy nuclei}, Phys. Rev. C {\bf94} 024320 (2016).
\href {https://doi.org/10.1103/PhysRevC.94.024320}
  {https://doi.org/10.1103/PhysRevC.94.024320}

\bibitem{Saxena:2024vtg} G.~Saxena, P.K. Sharma, P.~Saxena, {A global study of $\alpha $-clusters decay in heavy and superheavy nuclei with half-life and preformation factor}, Eur. Phys. J. A {\bf60} 50 (2024).
\href {https://doi.org/10.1140/epja/s10050-024-01259-w}
  {https://doi.org/10.1140/epja/s10050-024-01259-w}

\bibitem{Wan:2021wny} N.~Wan, J.~Fan, {Systematical calculations on {\ensuremath{\alpha}}-cluster
  preformation factors and decay half-lives of light nuclei near the recently observed {\ensuremath{\alpha}} emitters \ce{^{108}Xe} and \ce{^{104}Te}}, Phys. Rev. C {\bf104} 064320 (2021).
\href {https://doi.org/10.1103/PhysRevC.104.064320}{https://doi.org/10.1103/PhysRevC.104.064320}  

\end{thebibliography}
\end{document}